\documentclass[twocolumn,pra,floatfix]{revtex4-2}

\usepackage[colorlinks,pdfusetitle,urlcolor=blue,citecolor=blue,linkcolor=blue,bookmarksnumbered,plainpages=false]{hyperref}
\usepackage{graphicx}
\usepackage{bm}
\usepackage{amsmath}
\usepackage{amssymb}
\usepackage{braket}
\usepackage{leftidx}
\usepackage{placeins}

\renewcommand{\vec}[1]{\bm{#1}}

\newcommand{\ten}[1]{\mathbb{#1}}

\newcommand{\G}{\ten{G}}
\newcommand{\wkn}{\omega_{kn}}
\newcommand{\wmp}{\omega_{mp}}

\newcommand{\tr}{\operatorname{Tr}}

\newcommand{\diff}{\mathrm{d}}
\newcommand{\dif}{\mathrm{d}}
\newcommand{\mi}{\mathrm{i}}

\newcommand{\im}{\mathrm{Im}}

\newcommand{\s}{\hat S ^{(2)}}
\newcommand{\ra}{ \vec{r}_a}
\newcommand{\rb}{ \vec{r}_b}
\newcommand{\ta}{ t_a}
\newcommand{\tb}{ t_b}
\newcommand{\0}{\lbrace 0\rbrace}
\newcommand{\re}{\mathrm{Re}}
\newcommand{\rnr}{r_\mathrm{NR}}
\newcommand{\e}{ \mathrm{e}}
\newcommand{\iso}{\text{(iso)}}

\begin{document}

\bibliographystyle{apsrev4-2}
\title{Purcell modification of Auger and interatomic Coulombic decay}

\author{Janine Franz}
\email{janine.franz@physik.uni-freiburg.de} 
\affiliation{Physikalisches Institut, Albert-Ludwigs-Universit\"at
Freiburg, Hermann-Herder-Str. 3, 79104 Freiburg, Germany}

\author{Stefan Yoshi Buhmann}\email{stefan.buhmann@uni-kassel.de}
\affiliation{Institut für Physik, Universit\"at Kassel, Heinrich-Plett-Straße 40, D-34132 Kassel, Germany}

\date{\today}

\begin{abstract}
An excited two-atom system can decay via different competing relaxation processes. If the excess energy is sufficiently high the system may not only relax via spontaneous emission but can also undergo interatomic Coulombic decay (ICD) or even Auger decay. We provide analytical expressions for the rates by including them into the same quantum optical framework on the basis of macroscopic quantum electrodynamics.
By comparing the rates in free space we derive the atomic properties determining which decay channel dominates the relaxation. 
We show that by modifying the excitation propagation of the respective process via macroscopic bodies, in the spirit of the Purcell effect, one can control the ratio between the two dominating decay rates. We can relate the magnitude of the effect to characteristic length scales of each process, analyse the impact of a simple close-by surface onto a general two-atom system in detail and discuss the effect of a cavity onto the decay rates. We finally apply our theory to the example of a doubly excited HeNe-dimer.
\end{abstract}
%

\maketitle


\section{Introduction}
The simplest decay process an excited atom may undergo is spontaneous decay. 
Its modification by a perfectly reflecting plate was first considered theoretically in 1946 by Purcell and experimentally measured in 1970 \citep{Purcell1946}. Since then more complex macroscopic bodies have been considered as means to control the decay, including dielectric surfaces and cavities \cite{Agarwal1975,Dung2000,Scheel2009,Barrett2020}. 
In a similar manner, environments can be designed to modify other atomic processes. A prominent example is (F\"orster) resonant energy transfer (RET), where an initial excitation is transferred from a donor atom to another acceptor atom, leaving the system again in an excited state. 
The modification of RET has been extensively studied in theory and experiment  
and it could be shown in the framework of macroscopic quantum electrodynamics that f.e.~a close-by surface can support the transfer from donor to acceptor via surface waves \citep{Cortes2018, Ribeiro2018, Jones2019}. 
\\ \indent
Inspired by the success of these methods for the control of spontaneous decay and RET, we consider the decay of highly excited two-atom systems in a similar manner, where interatomic Coulombic decay (ICD) and Auger decay appear as competing relaxation channels.
Similar to RET, in ICD an excited donor atom relaxes and the energy is subsequently absorbed by an acceptor atom. However, in ICD the absorbed energy is sufficient to ionise the acceptor. ICD was only predicted theoretically in 1997 and experimentally measured almost ten years later \citep{ICDCederbaum1997, Marburger2003} and has received much attention since then \citep{Jahnke2020}.
Auger decay on the other hand is a radiationless autoionisation process where the excited electron makes a downward transition, while a second electron of the same atom is emitted into continuum and has been since its theoretical prediction in 1925 subject to theoretical as well as experimental studies for a long time \citep{Auger1923, Ku2019, Armen2000}.
\\ \indent
From their close relation one can already assume that ICD can be modified similairly as RET, this was shown in theory
\citep{Dung2002, Hemmerich2018}. 
On the other hand, Auger decay was only recently included into and studied in the same framework \citep{Franz2021}.
Both decays of interest involve the exchange of an excitation between two electrons in the XUV to x-ray regime. Together with the recent advances of x-ray sources, the field of x-ray optics, such as x-ray cavities, experiences rapid development \citep{Adams2013, Heeg2013}. We therefore lay the theoretical groundwork to treat these processes in the presence of such environments in an analytical manner and joint framework and predict their possible modifications. 
Once the Auger decay channel is energetically accessible, it typically dominates the overall decay and we hence focus our discussion on the possible enhancement of ICD over Auger decay.
\\ \indent
In Sect.~\ref{sec:mQED} of this article we introduce the basics of macroscopic quantum electrodynamics, including the classical Green's tensor and its properties. 
We then derive analytical expressions for ICD and Auger rates as two special cases of electron--electron scattering and transfer the process over to a quantum optical description in dipole approximation (Sect.~\ref{sec:scatt}). 
In the Sect.~\ref{sec:channels} we compare the (single-photon-exchange) relaxation channels in a highly excited two-atom system via the derived analytical expressions.
Section~\ref{sec:surface} is devoted to studying the influence of a simple close-by dielectric surface onto the ICD and Auger decay rate for general two-atom systems. 
Section~\ref{sec:cavity} offers a discussion on the expected maximum enhancement for each rate in a cavity. 
We finally apply our results to the example of a doubly excited HeNe-dimer in Sect.~\ref{sec:HeNe}.
\section{Macroscopic quantum electrodynamics and classical Green's tensor}
\label{sec:mQED}
Macroscopic quantum electrodynamics (mQED) describes the excitation of a polarizable medium and of the electromagnetic field as one common excitation. The annihilation and creation operators $\hat{\vec{f}}^{(\dagger)} $ for this combined body--field system fulfil the joint bosonic commutation relations, from which one can directly derive the expectation values of terms quadratic in the fundamental operators \citep{DF1, Buhmann2007}
\begin{align}
\bra{\0}
\hat{\vec{f}}(\vec{r},\omega) 
\hat{\vec{f}}^\dagger(\vec{r}',\omega) 
\ket{\0}
&=
 \bm{\delta}(\vec{r} - \vec{r}') \delta(\omega-\omega')
\nonumber
\\
\bra{\0}
\hat{\vec{f}}^{(\dagger)}(\vec{r},\omega) 
\hat{\vec{f}}^{(\dagger)}(\vec{r}',\omega) 
\ket{\0}
&=\bm{ 0 }
 \label{QED:ff}
\end{align}
The field can be coupled to the atomic system via: 
\begin{equation}
\hat{V}(t)  = 
				\int \!\! \dif V  
				\left[
				\hat \rho(\vec{r},t) \hat{\phi}(\vec{r},t)
				- 
				\hat{\vec{j}} (\vec{r},t) \!\cdot\!\hat{\vec{A}}(\vec{r},t) \right]
\label{QED:V}
\end{equation}
with $\hat{\rho}$, $\hat{\vec{j}}$ being the atomic charge density and current density operator, respectively.
The scalar and vector potentials of the field operators can be expanded in terms of these operators and read in Coulomb gauge:
\begin{align}
\nabla \phi(\bm r )
&= 
-
\int \!\! \dif \omega 
\underline{\hat{\vec{E}}}^\parallel (\vec{r},\omega) 
- \text{h.c}
\nonumber
\\
&=
-\int \!\! \dif \omega \!\!
\int \!\! \dif^3 r'
\mi
\frac{\omega}{c^2} 
\sqrt{\frac{\hbar}{\pi  \epsilon_0} \im \varepsilon(\vec{r}',\omega)} 
		\nonumber
\\ & \qquad\qquad  \times
{^\parallel} \ten{G}
(\vec{r},\vec{r'},\omega)\! \cdot \!\hat{\vec{f}}(\vec{r}',\omega)
- \text{h.c}
\label{QED:EofG}
\\
\bm{\hat{A}}(\bm r )
&= 
\int \!\! \dif \omega \!\!
\int \!\! \dif^3 r'
\frac{\omega}{c^2} 
\sqrt{	\frac{\hbar}{\pi  \epsilon_0} 
		\im \varepsilon(\vec{r}',\omega)} 
		\nonumber
\\ & \qquad\qquad  \times
{^\perp} \ten{G}
(\vec{r},\vec{r'},\omega)
\! \cdot \!
\hat{\vec{f}}(\vec{r}',\omega) 
+ \text{h.c.}
\label{QED:AofG}
\end{align}
The appropriate Green's tensor $\G(\bm r, \bm r', \omega)$
describes the propagation of the excitations throughout the environment, fulfilling its respective Helmholtz equation and boundary conditions at interfaces: 
\begin{align}
\left[
\bm \nabla \times \frac{1}{\mu(\bm r, \omega)} \bm \nabla 
\times
-
\frac{\omega^2}{c^2}\varepsilon(\bm r, \omega)
\right]\G (\bm r, \bm r',\omega) 
= \bm \delta (\bm r - \bm r'),
\end{align}
where we will restrict ourselves in this work to non-magnetic, homogeneous bodies: $\mu = 1$ and $\varepsilon(\bm r, \omega) = \varepsilon_i(\omega)$, for $\bm r \in V_i$, where $V_i$ is the volume of body $i$.
 Some useful properties can be generally derived for the Green's tensor in reciprocal media:
\begin{align}
\G ^T (\rb,\ra,\omega)&= \G( \ra, \rb ,\omega)  
\label{QED:onsager}
\\
\G^* (\rb,\ra, \omega) &= \G(\rb,\ra,-\omega^*)  
\end{align}
and the integral relation:
\begin{align}
\int \! \dif^3 r' 
\frac{\omega^2}{c^2} 
\im \varepsilon(\vec{r}',\omega ) 
\ten{G} (\rb,\vec{r}',\omega) \cdot \G ^{*T}(\ra,\vec{r}',\omega) 
\nonumber \\
= \im \G(\rb, \ra,\omega)
\label{QED:intergralrelation}
\end{align}
In free space ($\varepsilon(\bm r, \omega) = 1$) the Green's tensor $\G^{(0)}$ is given by: 
\begin{align}
\G^{(0)} 
(\ra,\rb, \omega)
&=
\frac{1}{4\pi} \left[ \ten{I} + \frac{c^2}{\omega^2} \bm \nabla_a \bm \nabla_a
\right] \frac{e^{\,i \omega |r_{ab}|/c}}{r_{ab}}
\nonumber
\\
&=
\frac{c^2}{3 \omega^2}\bm \delta (\bm r_{ab} ) 
- 
\frac{\e^{\mi \omega r_{ab}/c}}{4 \pi \omega^2 r_{ab}^3}
\nonumber
\\
&\qquad
\times
\bigg\lbrace 
	\left[
		1 - \mi \frac{\omega r_{ab} }{c} - \frac{\omega^2r_{ab}^2}{c^2} 
	\right] \ten{I}
\\
&\qquad\quad
- 	\left[ 
		3 - 3\mi \frac{\omega r_{ab}}{c} -\frac{\omega^2r_{ab}^2}{c^2} 
	\right] \bm e_{ab} \otimes \bm e_{ab} 
\bigg\rbrace 
\nonumber
\label{G0}
\end{align}
where $\bm r_{ab} = \bm r_b - \bm r_a$ and $\bm e_{ab} =\bm r_{ab} / r_{ab}$.
In the nonretarded limit $\omega r_{ab} /c \ll 1$ the Green's tensor can be approximated by: 
\begin{align}
\G^{(0)}_{\text{NR}}
(\rb,\ra, \omega)
=
 - \frac{c^2}{4\pi \omega^2 r_{ab}^3} \left( \ten{I} - 3 \bm e_{ab} \otimes \bm e_{ab}  \right)
\end{align}
which is real and diverges for $r_{ab} \rightarrow 0$. The imaginary part of the Green's tensor on the other hand stays finite in this limit: 
\begin{align}
\im \G^{(0)} 
(\bm r,\bm r , \omega)
&= 
\frac{\omega}{6 \pi c} \ten{I}
\end{align}
When adding surfaces one can obtain the full propagator $\G$ by adding the appropriate scattering Green's tensor $\G^{(1)}$ to the free--space solution $\G^{(0)}$. The scattering Green's tensor is known for several different geometries. 
For a homogeneous dielectric surface the scattering Green's tensor can be analytically calculated via an image dipole approach if the source-surface distance is in the nonretarded $\omega\Delta r /c \ll 1$ regime: 
\begin{align}
\begin{split}
\G^{(1)}_\text{surface}
(\rb,\ra, \omega)
&=
- \frac{ r_\mathrm{NR} c^2}{4\pi \omega^2 \bar{ r }_{ab}^3} ( \ten{I} - 3  \bar{\bm e}_{ab} \otimes \bar{\bm e}_{ab} )\cdot \ten{M}
\\
\ten{M} 
&= \ten{I} - 2 \bm e_{\Delta r} \otimes \bm e_{\Delta r} 
\\
\bar{\bm  r }_{ab}
&= \ten{M} \cdot \bm r_{ab} - 2 \Delta \bm r \mathrm{,}
\quad 
\bar{\bm e}_{ab} = 
\bar{\bm  r }_{ab}/{\bar r_{ab}}
\end{split}
\label{QED:G1surf}
\end{align}
where $r_\mathrm{NR} = ( \varepsilon - 1)/(\varepsilon+1)$ is the reflection coefficient in the nonretarded limit and $\varepsilon$ is the complex-valued relative permittivity of the material. 
Inside a cavity one can often work in the opposite limit, the retarded regime of large source-surface distances. The exact expression of the scattering Green's tensor depends on the geometry. For a spherical cavity with radius $R$ in the retarded regime of $\omega R/ c \gg 1$ and infinitely thick walls, we find for source and absorption in the center of the cavity:
\begin{align}
\begin{split}
\G^{(1)}_\text{cavity}
(0,0, \omega)
&= -e^{\mi k_0 R} \frac{  k_0 R (n^2- n ) +\mi \left(n^2-1\right) }
{D(k_0 R)}
\\
D (k_0 R)
&=
6 \pi  
\left[
(\mi k_0 R  (n-n^2)+n^2-1 ) \cos (k_0 R)
\right.
\\
&\qquad\qquad \left.
+
\mi k_0 R n^2 \exp (-\mi  k_0 R)
\right],
\end{split}
\end{align}
where $k_0 = \omega/c$. This is a complex oscillating function, which describes propagating electromagnetic waves.
Even the mediation by a single additional atom can be introduced via the scattering Green's tensor \citep{Bennett2019}:
\begin{align}
&\G^{(1)}_\mathrm{atom}(\bm r_b, \bm r_a,\omega)
\nonumber
\\
&\qquad\quad= 
\mu_0 \omega^2 \G^{(0)} ( \bm r_b, \bm r, \omega) \cdot \bm{\alpha} ( \omega) \cdot \G^{(0)} (\bm r,\bm r_a,\omega),
\label{QED:G1med}
\end{align}
where $\bm{\alpha}(\omega)$ is the polarisability tensor of the mediating atom and $\bm r$ its position. 
\section{Electron--electron scattering}
\label{sec:scatt}
\begin{figure} 
\includegraphics[width = 0.6 \linewidth]{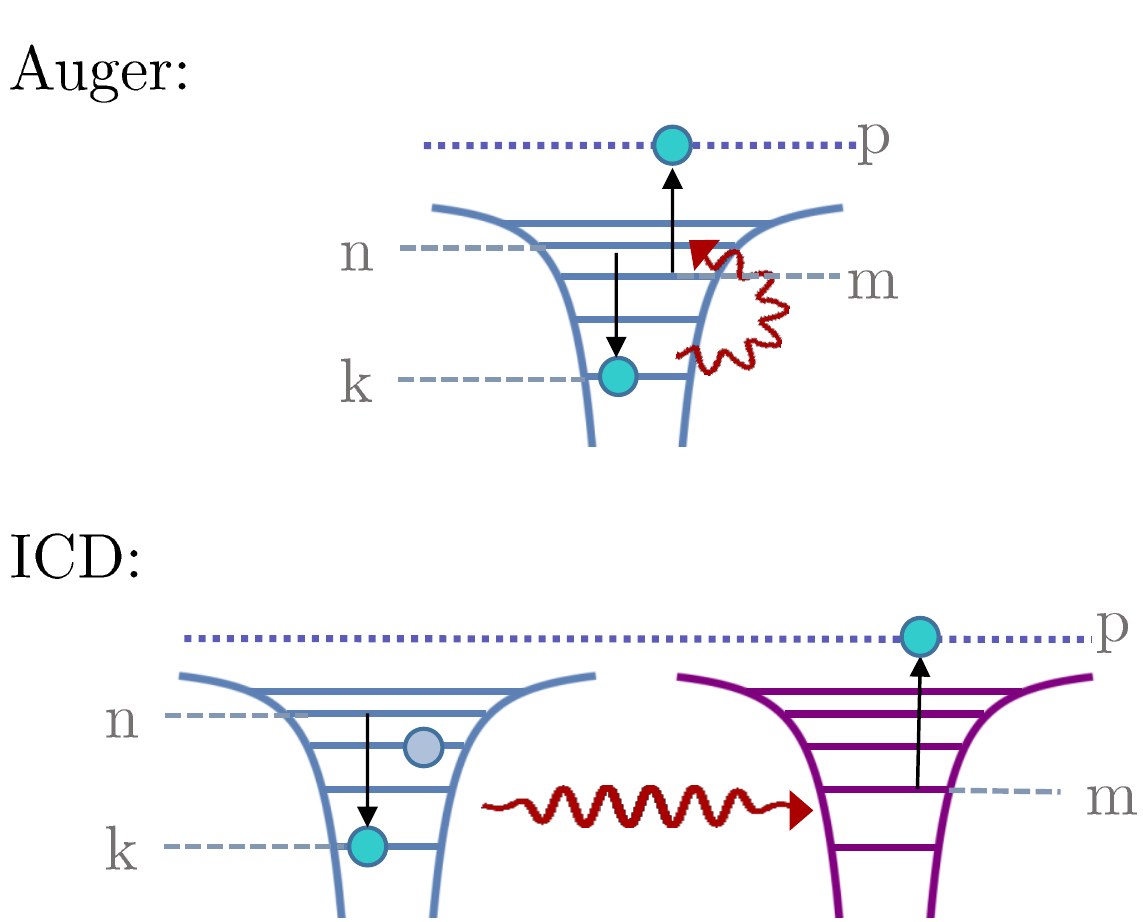}
\caption{Auger decay and ICD, schematically, including the chosen energy level labels. The initial state of the initially excited atom is the same for both processes.}
\label{fig:schemes}
\end{figure}
\begin{figure} 
\includegraphics[width = \linewidth]{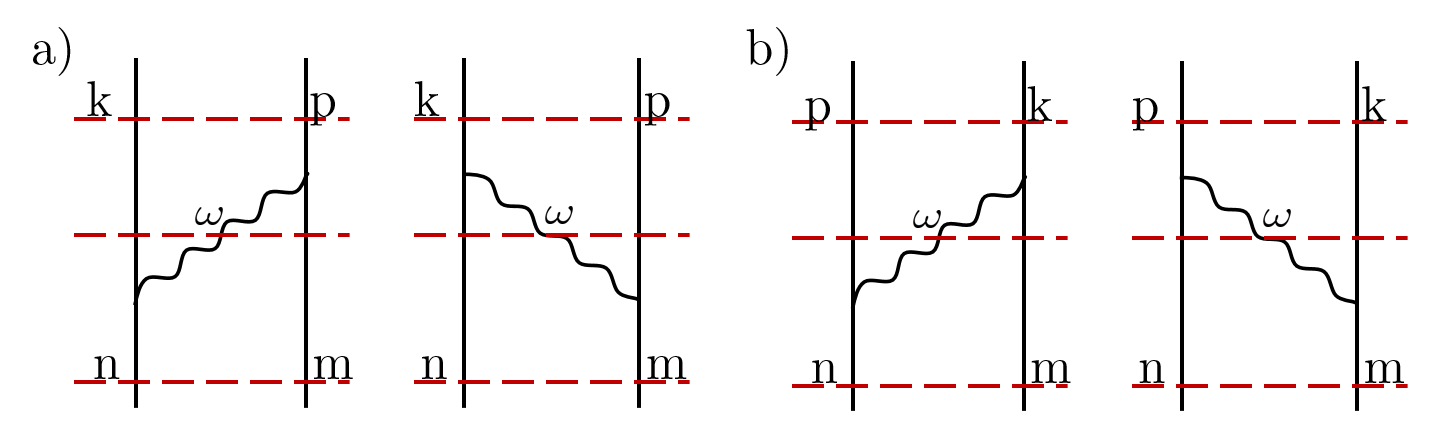}
\caption{Feynman diagrams for the Auger decay with the initial state of the two electrons being $\ket{n,m}$ and the final state $\ket{k,p}$. a) shows the direct terms in the two-electron interaction and b) the exchange terms. In both cases there are two different channels with different intermediate states (indicated by the red dashed line in the middle of each diagram). }
\label{fig:feynman}
\end{figure}
We derive the Auger decay and ICD rate as special cases of the same general process, an electron--electron scattering process, where the electrons interact with each other in second order via the electromagnetic field. In general a process rate $\Gamma$ can be expressed via the scattering matrix $\hat S(t)$:
\begin{equation}
\Gamma = 	\sum_f
			\frac{\partial}{\partial t} 
			| 
			\bra{f} 
			\hat{S}(t) 
			\ket{i} 
			|^2
			\label{rate}
\end{equation}
where $\ket{i}$, $\ket{f}$ are the initial and final states, respectively. 
Initially the two involved electrons are bound while the electromagnetic field is in its vacuum state $\ket{i} =\ket{\0} \ket{n,m}$, while in the final state $\ket{f} = \ket{\0} \ket{k,p}$ one electron is in the continuum state $\ket{p}$ while the other filled an initial vacancy $\ket{k}$ and again the electromagnetic field is without excitations. The two processes of interest including their level-labels are shown schematically in Fig.~\ref{fig:schemes}. 
The second order scattering operator $\hat S^{(2)}(t)$ is given by \citep{Chattarji1976}
\begin{equation}
\s(t) = 
			- \frac{1}{\hbar^2} 
			\int_{\text{-}\infty}^t \!\!\!\!\! \dif t_a  \!\!
			\int_{\text{-}\infty}^{t_a} \!\!\!\!\!  \dif t_b 
			\,
			\hat{V}(t_a) \hat{V}(t_b)
\end{equation}
where $\hat{V}(t)$ is the interaction Hamiltonian given bei Eq.~\eqref{QED:V}.
The scattering matrix then reads: 
\begin{align}
\s (t) 
&=
-\frac{1}{\hbar^2} 
[			
\hat{A}(t) + \hat{B}(t) + \hat{C}(t)],
			\label{S2}
\end{align}
with three contributions, that will be treated separately:
\begin{subequations}
\begin{align}
\hat{A}(t) 
&=\int_{\text{-}\infty}^t \!\!\!\!\!\! \dif t_a  \!\!
			\int_{\text{-}\infty}^{t_a} \!\!\!\!\!\!  \dif t_b \!\!
			\int \! \! \dif^3 r_a  \!\!
			\int \!\!\dif^3 r_b 
			\,\,
					\hat \rho_a 	\hat \phi _a 
					\hat \rho _b 	\hat \phi_b 
\\
\hat{B}(t) 
&=\int_{\text{-}\infty}^t \!\!\!\!\!\! \dif t_a  \!\!
			\int_{\text{-}\infty}^{t_a} \!\!\!\!\!\!  \dif t_b \!\!
			\int \! \! \dif^3 r_a  \!\!
			\int \!\!\dif^3 r_b 
			\,\,
					\hat {\vec{j}}_a \!\cdot\!
					\hat {\vec{A}}_a  \,
					\hat {\vec{j}}_b  \!\cdot \!
					\hat {\vec{A}}_b 
\\
\hat{C}(t) 
&= \!-\!\! \int_{\text{-}\infty}^t \!\!\!\!\!\! \dif t_a  \!\!
			\int_{\text{-}\infty}^{t_a} \!\!\!\!\!\!  \dif t_b \!\!
			\int \! \! \dif^3 r_a  \!\!
			\int \!\!\dif^3 r_b 
\nonumber
\\
&\qquad\qquad\qquad
			\left[
					\hat \rho_a 	\hat \phi _a   
					\hat {\vec{j}}_b \! \cdot \!
					\hat {\vec{A}}_b 
					+  
					\hat {\vec{j}}_a \!\cdot \!
					\hat {\vec{A}}_a  
					\hat \rho _b 		\hat \phi_b \right],
\end{align}
\end{subequations}
where we have used the shorthand notation $\hat{O}_a =\hat{O}(\vec{r}_a, t_a)$ for any operator $\hat O$.
Auger decay as well as ICD can be visualized as an exchange of a photon between the participating electrons, see Fig.~\ref{fig:feynman}. 
The process has therefore four different interfering decay channels, two of which result from the indistinguishability of the involved electrons (i.e. exchange terms).
The exchange term as well as the direct term can again be divided into two different diagrams with different intermediate states, where one of each describes a virtual photon exchange.
Since the electrons are fermions the exchange term gains a sign change: 
\begin{align}
\bra{f}  \s(t) \ket{i}
&=
\bra{k,p} \bra{\0} \s(t) \ket{\0} \ket{n,m} 
\nonumber 
\\
&= 
\underbrace{
		S^{(2)}_{n \rightarrow k}(t)}_{\text{direct term}} 
		- 
\underbrace{S^{(2)}_{m \rightarrow k}(t)}_{\text{exchange term}}
\label{s2element}
\end{align}
Let us consider the direct term in equation \eqref{s2element}, where one electron goes from state $\ket{n}$ to the energetically lower final state $\ket{k}$ while the other one changes from the bound state $\ket{m}$ to the continuum state $\ket{p}$. The exchange term will follow directly.
The second-order scattering matrix~\eqref{S2} consists of three different contributions, which can be treated in a very similar manner. We show the derivation for the first contribution $\hat{A}(t)$ of the direct term: 
\begin{align}
\braket{f| \hat{A}(t) |i}
&=
\int_{\text{-}\infty}^t \!\!\!\!\!\! \dif t_a  \!\!
\int_{\text{-}\infty}^{t_a} \!\!\!\!\!\!  \dif t_b \!\!
\int \! \! \dif^3 r_a  \!\!
\int \!\!\dif^3 r_b  \,\, 
\nonumber 
\\ 
& \qquad\qquad
\Big\lbrace
	\rho_{nk}^{(a)}
	\bra{\0} \hat \phi _a  \hat \phi_b \ket{\0} 
	\rho _{mp}^{(b)}
\nonumber 
\\ 
& \qquad\qquad\quad
+	\rho _{mp}^{(a)}
	\bra{\0} \hat \phi _a  \hat \phi_b \ket{\0} 
	\rho_{nk}^{(b)} 
	\Big\rbrace
\label{[A]}
\end{align}
where $\rho_{nk}^{(a)} = \braket{k|\rho (\ra,\ta) |n}$.
Since the interaction energy of the two electrons is small, we may regard the initial states of the electrons as stationary and we can explicitly use the time dependence of the transition charge as well as the transition current density:
\begin{align}
\begin{split}
\rho_{nk}(\bm r_a,t)
&=
e^{\mi \frac{(E_k - E_n)}{\hbar} t}	
	\rho_{nk}^{(a)}
=
e^{\mi \omega_{nk} t}	
	\rho_{nk}^{(a)}
\\
\vec{j}_{nk}(\vec{r}_a,t)  
&= e^{   \mi \omega_{nk} t}  \vec{j}_{nk}^{(a)},
\end{split}
\label{j(t)}
\end{align}
where we introduced the shorthand notation $O_{nk}^{(a)} = O_{nk} ( \bm r_a)$ for the time-independent transition element of an operator $\hat{O}$.
The continuity equation then yields:
\begin{align}
\rho_{nk}^{(a)}
&= \frac{\mi}{\omega_{nk}} \vec{\nabla} \cdot \vec{j}_{nk}^{(a)}
\label{conteq}
\end{align}
To simplify the following calculation we will already consider the different signs of the transition frequencies. For the downward transition from $\ket{n}$ to $\ket{k}$ one finds $E_k - E_n < 0$, we therefore define the positive frequency $\omega_{kn} = (E_n - E_k) / \hbar = - \omega_{nk} >0$. In case of the second transition ($\ket{m} \rightarrow \ket{p}$) the transition energy is positive ($\omega_{mp} = (E_p - E_m) / \hbar > 0 $). 
Combining Eqs.\,\eqref{[A]}~--~\eqref{conteq} we obtain:
\begin{align}
&\braket{f|\hat{A}(t) |i} 
=	
\int_{\text{-}\infty}^t \!\!\!\!\!\! \dif t_a  \!\!
\int_{\text{-}\infty}^{t_a} \!\!\!\!\!\!  \dif t_b \!\!
\int \! \! \dif^3 r_a  \!\!
\int \!\!\dif^3 r_b  \,\, 
	\frac{1}{\omega_{kn} \omega_{mp}}
\nonumber
\\
&\quad
\times
\bigg\lbrace
	 \vec{\nabla}\! \cdot\! \vec{j}_{nk}(\ra ,\ta)
	\bra{\0} \hat \phi _a  \hat \phi_b \ket{\0} 
	 \vec{\nabla}\! \cdot\! \vec{j}_{mp}(\rb ,\tb)
\nonumber \\
&\quad 
	\qquad\quad	
	+
	 \vec{\nabla} \!\cdot\!  \vec{j}_{mp}(\ra ,\ta)
	\bra{\0} \hat \phi _a  \hat \phi_b \ket{\0} 
	 \vec{\nabla} \! \cdot \! \vec{j}_{nk}(\rb ,\tb)
\bigg\rbrace
\nonumber
\\
&=	
\int_{\text{-}\infty}^t \!\!\!\!\!\! \dif t_a  \!\!
\int_{\text{-}\infty}^{t_a} \!\!\!\!\!\!  \dif t_b \!\!
\int \! \! \dif^3 r_a  \!\!
\int \!\!\dif^3 r_b  \,\,	\frac{1}{\omega_{kn} \omega_{mp}}
\nonumber
\\
& \qquad
\times
\bigg\lbrace
	\vec{j}_{nk}(\ra ,\ta)
	\!\cdot\!
	\bra{\0} 
	\hat{\vec{E}}^\parallel_a 
	\hat{\vec{E}}^\parallel_b 
	\ket{\0} 
	\!\cdot\!
	\vec{j}_{mp}(\rb ,\tb)
\nonumber \\
&\qquad \quad \quad 
+
	 \vec{j}_{mp}(\ra ,\ta)
	 \!\cdot\!
	\bra{\0} 
	\hat{\vec{E}}^\parallel_a 
	\hat{\vec{E}}^\parallel_b 
	\ket{\0} 
	\! \cdot\!
	 \vec{j}_{nk}(\rb ,\tb)
\bigg\rbrace
\end{align}
where we integrated by parts.
By using the Fourier-transform of $\hat{\vec{E}}^\parallel(\bm r ,t)$ and  expand $\hat{\vec{E}}(\bm r, \omega)$ in terms of the Green's tensor, see Eq.\,\eqref{QED:EofG}, we can evaluate the expectation value in the electromagnetic vacuum by means of its annihilation and creation operators, see Eq.\,\eqref{QED:ff} and use the integral relation Eq.\,\eqref{QED:intergralrelation} to obtain the first contribution of the direct term:
\begin{align}
\braket{f|\hat{A}(t) |i} 
&= 
\int_{\text{-}\infty}^t \!\!\!\!\!\! \dif t_a  \!\!
\int_{\text{-}\infty}^{t_a} \!\!\!\!\!\!  \dif t_b \!\!
\int \! \! \dif^3 r_a  \!\!
\int \!\!\dif^3 r_b\!\!
	\int_0^\infty \!\!\!\!\!\!\!\dif \omega
\frac{	e^{\mi \omega (t_b -t_a)} }
			{\omega_{kn} \omega_{mp}}
\frac{\hbar \omega^2}{\pi \epsilon_0 c^2}
\nonumber
\\
& \quad \times
\bigg\lbrace
\Big(
	e^{- \mi \omega_{kn} t_a }
	e^{\mi \omega_{mp} t_b }  
	+
	e^{\mi \omega_{mp} t_a }
	e^{- \mi \omega_{kn} t_b }
\Big)
\nonumber \\& \qquad \quad \times
		\vec{j}_{nk}^{(a)}
		\!\cdot\! 
		\im {^\parallel}\G^\parallel_{ab} (\omega)
		\!\cdot \!
		\vec{j}_{mp}^{(b)}
\bigg\rbrace
\end{align}
where we used that the term is integrated over $\vec{r}_b$ as well as $\vec{r}_a$ together with the Onsager reciprocity \eqref{QED:onsager} and the known time dependence of the charge currents \eqref{j(t)} and we introduced the shorthand notation $\G_{ab} (\omega) =\G(\bm r_a, \bm r_b \omega)$.
To carry out the time integration we need to introduce a switching parameter $\epsilon > 0$ into the interaction potential $\hat{V}(t) \rightarrow \hat{\tilde{V}}(t) = \mathrm{e}^{\epsilon t} \hat{V}(t) $. This parameter explicitly ensures the premise that the initial state $\ket{i}$ can be thought of as unperturbed for $t \rightarrow - \infty$. The time integration then yields:
\begin{align}
\begin{split}
&
\int_{\text{-}\infty}^t \!\!\!\!\!\! \diff \ta  \!\!
\int_{\text{-}\infty}^{\ta} \!\!\!\!\!\! \diff \tb \, 
e^{\epsilon (t_a + t_b) }
e^{\mi \omega (\tb - \ta )} 
\Big\lbrace 
	e^{- \mi \omega_{kn} t_a }
	e^{\mi \omega_{mp} t_b }
 \\ & 	\qquad\qquad\qquad\qquad\qquad\qquad\qquad
	+
	e^{\mi \omega_{mp} t_a }
	e^{- \mi \omega_{kn} t_b }
\Big\rbrace
\\
&=
\lim_{\epsilon \rightarrow 0^+}
f_\epsilon (t) 
\Bigg\lbrace
\frac{1}{\mi (\omega + \wmp - \mi \epsilon)  }
+
\frac{1}{\mi (\omega - \wkn - \mi \epsilon)  }
\Bigg\rbrace
\\
& \quad \text{with: } \quad 
f_\epsilon(t) = 
\frac{ e^{\mi (\wmp -\wkn) t } e^{\epsilon t }}
		{\epsilon + \mi (\wmp - \wkn)}
\end{split}
\end{align}
Carrying out the frequency integration gives:
\begin{align}
\braket{f|\hat{A}(t)|i}
&= 
\int \!\! \dif^3 r_a  \!\!
\int \! \!\dif^3 r_b  \lim_{\epsilon \rightarrow 0^+} 
\frac{\hbar f_\epsilon(t) }{\pi \varepsilon_0 c^2}
\!\!
\nonumber
\\
&\qquad
	\times
	\Bigg\lbrace
	- \mi \pi
	\frac{\wkn}{\wmp}
	\bm j^{(a)}_{nk}
	\! \cdot \!
	{^\parallel}\G_{ab}^\parallel (\wkn)
	\! \cdot \!
	\bm j^{(b)}_{mp}
	\nonumber
	\\
	& \qquad\quad
	+ \mi
\int \!\! \diff \omega \,
	\frac{\omega^2}{\wmp\wkn}
	\bm j^{(a)}_{nk}
	\cdot
	\frac{\im {^\parallel}\G_{ab}^\parallel (\omega)}
	{\omega+\wkn}
	\cdot
	\bm j^{(b)}_{mp}
	\nonumber
	\\
	& \qquad\quad
	- \mi
\int \!\! \diff \omega \,
	\frac{\omega^2}{\wmp\wkn}
	\bm j^{(a)}_{nk}
	\cdot
	\frac{\im {^\parallel}\G_{ab}^\parallel (\omega)}
	{\omega+\wmp}
	\cdot
	\bm j^{(b)}_{mp}
	\Bigg\rbrace
\nonumber 
\\
&= \lim_{\epsilon \rightarrow 0^+} 
\frac{\hbar f_\epsilon(t) }{\mi \varepsilon_0 c^2}
 \!\!
\int \!\! \dif^3 r_a  \!\!
\int \! \!\dif^3 r_b \, 
	\bm j^{(a)}_{nk}
	\! \cdot \!
	{^\parallel}\G_{ab}^\parallel (\wkn)
	\! \cdot \!
	\bm j^{(b)}_{mp}
\end{align} 
where we already exploited the fact that the rate will be evaluated at $\wkn = \wmp $, see Eq.\,\eqref{gamma_scatt}.
By using the expansion of $\bm{\hat{A}}(\bm r , t)$ in terms of the Green's tensor, see Eq.~\eqref{QED:AofG}
the remaining two contributions $\hat{B}(t)$ and $\hat{C}(t)$ can be calculated in a very similar manner:
\begin{align}
\braket{f|\hat{B}(t)|i}
& = 
\lim_{\epsilon \rightarrow 0^+} 
\frac{\hbar f_\epsilon(t) }{\mi \varepsilon_0 c^2}
\!\!
\int \!\!\! \dif^3 r_a  \!\!\!
\int \!\! \!\dif^3 r_b \,
	\bm j^{(a)}_{nk}
	\! \cdot\!
	{^\perp}\G_{ab}^\perp (\wkn)
	\! \cdot\!
	\bm j^{(b)}_{mp}
\\
\braket{f|\hat{C}(t)|i} 
& = 
\lim_{\epsilon \rightarrow 0^+} 
\frac{\hbar f_\epsilon(t) }{\mi \varepsilon_0 c^2}
\int \!\! \dif^3 r_a  \!\!
\int \! \!\dif^3 r_b \,
	\bm j^{(a)}_{nk}
\nonumber
\\
& \qquad\qquad\qquad
	\! \cdot\!
\Big[	
	{^\perp}\G_{ab}^\parallel (\wkn)
	+{^\perp}\G_{ab}^\parallel (\wkn) 
	\Big]
	\! \cdot\!
	\bm j^{(b)}_{mp}
\end{align}
With $\G = {^\parallel}\G^\parallel 
	+{^\perp}\G^\parallel + {^\parallel}\G^\perp + {^\perp}\G^\perp$ we finally obtain for the direct term: 
\begin{align}
S_{n\rightarrow k} 
& =
\lim_{\epsilon \rightarrow 0^+ } f_\epsilon(t) \, V_{n \rightarrow k}
\\
\text{with:}
& \nonumber
\\
V_{n \rightarrow k}
& = 
\frac{\mi }{ \hbar\varepsilon_0 c^2}
\!\!
\int \!\!\! \dif^3 r_a  \!\! 
\int \!\! \!\dif^3 r_b \,
	\bm j^{(a)}_{nk}
	\! \cdot\!
	\G_{ab}  (\wkn)
	\! \cdot\!
	\bm j^{(b)}_{mp}
\label{SinjGjdir}
\end{align}
where $V_{n\rightarrow k}$ is time-independent. The exchange term $S_{n\rightarrow p}$ can be obtained by switching the indices accordingly. This yields the same time dependence $f_\epsilon(t)$, since $\wmp - \wkn = \omega_{mk} - \omega_{pn}$.
The rate \eqref{rate} is then given by: 
\begin{align}
\Gamma 
&= \sum_f \frac{d}{dt} |\bra{f} \s(t) \ket{i}|^2
\nonumber
\\
&= \sum_f 
\frac{d}{dt}
\left| 
			\lim_{\epsilon \rightarrow 0^+}
			f_\epsilon(t)
			\right|^2
\left| V_{n\rightarrow k} -V_{n \rightarrow p} \right|^2
\nonumber
\\
&=
\sum_f 2 \pi \delta(\wmp -\wkn)
\left| V_{n\rightarrow k} -V_{n \rightarrow p} \right|^2
\nonumber
\\
&= 2 \pi 
\sum_{m,n}\int \!\! \diff \omega_p \rho(\omega_{mp}) 
\delta(\wmp -\wkn)
\left| V_{n\rightarrow k}- V_{n \rightarrow p} \right|^2
\nonumber
\\
&= 2 \pi 
\sum_{m,n}
\rho( \omega_{mp} ) 
\left| V_{n\rightarrow k} - V_{n \rightarrow p} \right|^2
\label{gamma_scatt}
\end{align}
where $\rho(\omega_{mp})$ is the density of final states of the continuum state at energy $\omega_{mp}$. 
Assuming that there exists only one initial vacancy, possible final states differ in their final vacancies in states $\ket{n}$ and $\ket{m}$. From this point on we will assume that each sum only involves degenerate final states, for simplicity. This formula is closely related to Fermi's golden rule. One can show that this is equivalent to the famous M\o ller formula when inserting the free space Green's tensor \eqref{G0} into \eqref{SinjGjdir}:
\begin{align}
V_{n\rightarrow k}
&= 
\frac{\mi }{ \hbar\varepsilon_0 c^2}
\!\!
\int \!\!\! \dif^3 r_a  \!\! 
\int \!\! \!\dif^3 r_b \,
	\bm j^{(a)}_{nk}
	\! \cdot\!
	\left[ \ten{I} + \frac{c^2}{\wkn^2} \bm \nabla_a \bm \nabla_a
\right] 
\nonumber 
\\
&\qquad\qquad\qquad\qquad\qquad\qquad
\times 
\frac{e^{\,i \wkn |r_{ab}|/c}}{4\pi r_{ab}}
	\! \cdot\!
	\bm j^{(b)}_{mp}
\nonumber
\\
&=
\frac{\mi }{ \hbar\varepsilon_0 c^2}
\!\!
\int \!\!\! \dif^3 r_a  \!\! 
\int \!\! \!\dif^3 r_b \,
\bigg[
\bm j^{(a)}_{nk} \cdot \bm j^{(b)}_{mp}
\nonumber
\\
&\qquad\qquad
-
\bm j^{(a)}_{nk} 
\cdot
\left( 
\bm \nabla_a \bm \nabla_b \frac{c^2 e^{\,i \wkn |r_{ab}|/c}}{4\pi \wkn^2 r_{ab}}
\right)
\cdot
\bm j^{(b)}_{mp}
\bigg]
\nonumber
\\
&=
\frac{\mi }{ \hbar\varepsilon_0 c^2}
\!\!
\int \!\!\! \dif^3 r_a  \!\! 
\int \!\! \!\dif^3 r_b \,
\frac{e^{\,i \wkn |r_{ab}|/c}}{4\pi r_{ab}}
\bigg[
\bm j^{(a)}_{nk} \cdot \bm j^{(b)}_{mp}
\nonumber
\\
&\qquad\qquad\qquad
- \frac{c^2}{\wkn^2}
\left(\bm \nabla_a \cdot
\bm j^{(a)}_{nk} 
\right)
\left(\bm \nabla_b 
\cdot
\bm j^{(b)}_{mp}\right)
\bigg]
\nonumber
\\
&=
\frac{\mi }{ \hbar\varepsilon_0 c^2}
\!\!
\int \!\!\! \dif^3 r_a  \!\! 
\int \!\! \!\dif^3 r_b \,
\frac{e^{\,i \wkn |r_{ab}|/c}}{4\pi r_{ab}}
\nonumber
\\
&\qquad\qquad\qquad\qquad
\times
\bigg[
\bm j^{(a)}_{nk} \cdot \bm j^{(b)}_{mp}
- 
c^2
\rho^{(a)}_{nk} 
\rho^{(b)}_{mp}
\bigg],
\end{align}
which is the M\o ller-formula for electron--electron scattering when plugged into \eqref{gamma_scatt} \cite{Chattarji1976}.
\subsection{Dipole approximation}
The calculation of the process rate has now boiled down to the calculation of the transition matrix elements:
\begin{align}
V_{n\rightarrow k} = 
\frac{\mi \mu_0}{ \hbar} 
\int \!\!\dif^3 r_a  \!\! 
\int \!\! \dif^3 r_b \,\,
	\bm j^{(a)}_{nk}
	\! \cdot\!
	\G_{ab}  (\wkn)
	\! \cdot\!
	\bm j^{(b)}_{mp}
\label{Vnk}
\end{align}
In the dipole approximation the transition charge currents can be expressed by transition dipole moments via the Thomas-Reiche-Kuhn sum rule:
\begin{align}
\bm j^{(a)}_{nk}
&= 
\braket{k| \hat{\bm j}(\bm r_a) |n}
\nonumber 
=
\sum_\alpha \frac{q_\alpha}{m_\alpha}  \braket{k| \hat{\bm p}_\alpha \delta(\bm r_a  -\hat{\bm R}) |n} 
\\
&= 
\wkn \bm d_{nk} \delta(\bm r_a - \bm R_a)
\end{align}
where $\bm p_\alpha$ is the momentum operator and $m_\alpha$ the mass of electron $\alpha$, $\hat{\bm R}$ is the center of mass position operator and $\bm R_a$ the position of the nucleus belonging to electron $\alpha$.
Introducing this into equation \eqref{Vnk} we find: 
\begin{align}
V_{n\rightarrow k} 
&= 
-\frac{\mi  \mu_0 \wkn^2}{\hbar} 
\bm d_{nk} \cdot \G_{ab}(\wkn) \cdot \bm d_{mp} 
\end{align}
It would also be possible to include the effect of electronic wavefunction overlap by taking Eq.~\eqref{Vnk} and apply the dipole approximation only to the addend involving the scattering part of the Green's tensor $\G = \G^{(0)} + \G^{(1)}$, then solving the integrations via usual methods of \textit{ab initio} quantum chemistry.
\subsection{Interatomic Coulombic decay}
In the interatomic Coulombic decay the electron of a donor atom transitions from state $\ket{n}$ to the lower vacant state $\ket{k}$, while the second electron belongs to an acceptor atom and transitions from state $\ket{m}$ to the continuum (state $\ket{p}$). Since the separation between the atoms is assumed to be sufficiently large so that orbital overlaps between donor and acceptor can be neglected, the exchange term of the process vanishes and we arrive at the ICD rate:
\begin{align}
\Gamma_{\text{ICD}} 
&= 
\frac{2 \pi \mu_0^2 }{\hbar^2} \sum_{m,n} 
\rho(\omega_p) \wkn^4 \big| \bm d_{nk} \cdot \G_{ab} \cdot \bm d_{mp}  \big|^2
\label{GammaICD1}
\end{align}
which can alternatively be derived from multipolar coupling in dipole approximation to the electromagnetic field and Fermi's golden rule \cite{Hemmerich2018}.
Assuming that the involved atoms are not aligned in any specific way to each other we may use the isotropic average: $
\bm d_{yx}\bm d_{xy} = 
\frac{1}{3}|\bm d_{xy}|^2 \ten{I}$, which gives:
\begin{align}
\Gamma^\iso_{\text{ICD}}  
 &= 
 \frac{2 \pi \mu_0^2 }{9 \hbar^2}  
\rho(\omega_p) \wkn^4 
|\bm d_{nk}|^2|\bm d_{mp}|^2
\nonumber
\\
&\qquad\qquad\qquad\qquad \times
 \tr \!\big[ \G_{ab}(\wkn) 
 \cdot \G^*_{ba}(\wkn) \big]
 \nonumber
 \\
&= 
2\pi
\gamma_{nk} \sigma_m(\wkn)
 \tr \!\big[ \G_{ab}(\wkn) 
 \cdot \G^*_{ba}(\wkn) \big],
 \label{GammaICDiso}
\end{align}
where we used that $\G_{ab}^T = \G_{ba}$ and that $\ket{p}$ is a continuum state, so we can relate the respective transition dipole moment to the photo ionisation cross section via: 
$\sigma_m(\wmp) = \rho(\omega_p) \frac{\pi \wmp}{3 \varepsilon_0 c \hbar} |\bm d_{mp}|^2$\cite{hilborn1982einstein}, similarly we replaced the transition dipole moment of the bound states with the respective spontaneous decay rate: $\gamma_{nk} = \frac{\mu_0 \wkn^3 }{3 \pi \hbar c} |\bm d_{kn}|^2$.
This is the final expression for isotropic ICD in an arbitrary environment.
To obtain the ICD rate in free space we use the free space Green's tensor \eqref{G0} in Eq.~\eqref{GammaICDiso}. This yields in the nonretarded limit $\omega  r_{ab} /c \ll 1$: 
\begin{align}
\Gamma_\mathrm{0,ICD}^\iso
&\approx
\frac{3 c^4 }{4 \wkn ^4 r_{ab}^6} \gamma_{nk} \sigma_m (\wkn)
\label{GammaICDiso0NR}
\end{align}
\subsection{Auger decay}
The Auger process proves to be more intricate, since the dipole approximation has to fail. The infinite loop-propagation given by $\G(\bm R, \bm R, \omega)$ can be regularized by reintroducing the size of the atom back into the calculation approximately in form of an electron cloud with Gaussian shape. This leads to a effective regularized Green's tensor that is the convolution of the original bulk Green's tensor $\G^{(0)}$ and the electron clouds \cite{Franz2021}. Originally this method was used to improve results for van-der-Waals and Casimir-Polder forces involving separation distances comparable to the size of the atoms or molecules \cite{Parsons2009,Fiedler2018,Das2020}. By this procedure we regain a finite result for the loop propagation:
\begin{align}
\tilde{\G}^{(0)}(\omega) \approx  - \frac{c^2}{24 \pi^{3/2} a^3 \omega^2} \ten{I}
\label{G0A}
\end{align}
where the Auger-radius $a$ is the size of the Gaussian describing the electron cloud distribution and is of the order of the vacancy orbital radius. However if the vacuum rate for a given Auger process is known, one can calculate the respective bulk Green's tensor $\tilde{\G}^{(0)}$ and use it to take additionally the impact of the scattering part $\G^{(1)}$ of Green's tensor onto the Auger rate into account. 
In the Auger process the exchange term is in general not negligible. The rate is therefore given by the sum of the absolute squares of the direct and exchange term, respectively, $\Gamma_\text{pure}$ as well as a term that results from the interference of both terms $ \Gamma_\text{intf}$: 
\begin{align}
\Gamma_\text{A} 
&= 2 \pi \sum_{m,n} \rho(\omega_p) 
\big\lbrace 
\overbrace{
|V_{n\rightarrow k}|^2 +  |V_{n\rightarrow p}|^2 }^{\Gamma_\text{pure}}
\nonumber
\\
&\qquad\qquad\qquad\qquad
+
\underbrace{
2 \text{Re}\!\left[V_{n\rightarrow k}V^*_{n\rightarrow p} \right]
}_{ \Gamma_\text{intf}}
\big\rbrace
\end{align}
In the previous case of ICD we considered two isotropically aligned atoms and used this to average over all possible orientations of the transition dipoles. In the case of the Auger process both transition dipoles arise in the same atom. The averaging is carried out by summing over all degenerate states. 
By applying the Wigner--Eckhart theorem and calculating the respective Clebsch--Gordon coefficients one can relate the dipole moments again to their absolute value squared in the pure part of the Auger rate:
\begin{align}
\Gamma_{\text{pure}}
 &= 
 18 \pi  c_{nkm}
\gamma_{nk} \sigma_{m} ( \omega_p)
 \tr \!\big[ \G (\wkn) \cdot \G^*  ( \wkn) \big]
\end{align}
where we have introduced once more the spontaneous decay rate as well as the photoionisation cross section and $c_{nkm}$ is a factor stemming from the Wigner--Eckhart theorem and the sum over $m$ and $n$ excludes now degeneracies. 
In the interference term the idea is similar, however in the general case it is only possible to separate the Green's tensors from the dipole moments in a less simple way: 
\begin{align}
\Gamma_{\text{intf}} 
 &= 
-
\frac{2 \pi \mu_0^2 }{\hbar^2} \sum_{m,n} \wkn^2 \omega_{km}^2
\ten{D}
::
\left[ \G (\wkn) \otimes \G ^*( \omega_{km} )\right] 
\end{align}
where $\ten{D} = \bm d_{nk} \otimes \bm d_{mp} \otimes \bm d_{pn} \otimes \bm d_{km}$ is a 4th-rank tensor and $\ten{A}::\ten{B} = \sum A_{ijkl} B_{ijkl}$ is the Frobenius inner product for 4th-rank tensors. Summing this again with help of the Wigner--Eckhart theorem over the degenerate states will give very few surviving elements.
Th expression can be further simplified when assuming, that the 
exchange term is either negligible (f.e. if $m \rightarrow k$ is dipole forbidden) or that the Auger decay is of the type XYY (f.e. KLL-decay) and that the transition dipole moments are isotropic by means of their degeneracies:
\begin{align}
\Gamma_\text{0,A}^\iso 
&= 
2 \pi \gamma_{yx} \sigma_{y}(\wkn) \tr\lbrace \tilde{\G}^{(0)}(\wkn)\cdot \tilde{\G}^{(0)*}(\wkn) \rbrace
\\
&\approx
 \frac{c^4 }{96 \pi ^2 a^6 \wkn^4} \gamma_{nk} \sigma_m(\wkn) 
\label{augerfreespace}
\end{align}
In this calculation we exploited that the rates of the total shell are independent of the chosen coupling scheme \cite{Crasemann1971}. It should be noted that the coupling-independence does not hold for Coster--Kronig transitions.

\section{Comparison of decay channels}
\label{sec:channels}
\begin{table*}
\centering
\renewcommand{\arraystretch}{3.5}
\begin{tabular}{|c|c|c|c|c|}
\hline 
\hspace{1.5cm}
  & $\Gamma_\mathrm{s,0}$ 
  & $\Gamma_\mathrm{A,0}$ 
  & $\Gamma_\mathrm{ICD}$ 
  &$ \Delta \Gamma_\mathrm{s} $
  \\ 
\hline 
$\Gamma_\mathrm{s,0}$ 
& 1 
&  
& 
& 
 \\ 
\hline 
$\Gamma_\mathrm{A,0}$ 
& 
$\dfrac{2 \pi^2}{3} \left( \dfrac{\lambda_{nk}}{4\pi^{3/2} a } \right)^4  \left( \dfrac{a_\sigma }{2 \pi^\frac{1}{2} a } \right)^2$ 
& 1 
& 
& 
   \\ 
\hline 
$\Gamma_\mathrm{ICD}$ 
& $ \dfrac{3}{4} \left( \dfrac{\lambda_{nk} }{2 \pi r}\right)^4    \left( \dfrac{a_\sigma }{r}\right)^2 $ 
& $\dfrac{9}{8 \pi^2} \left( \dfrac{2 \pi ^\frac{1}{2} a}{r} \right)^6 $
& 1 
& 
\\ 
\hline 
$\Delta \Gamma_\mathrm{s} $
& $3 \left( \dfrac{\lambda_{nk}}{2 \pi r}\right)^3   \dfrac{\im \tilde \alpha}{r^3} $ 
& $
12
\left( \dfrac{2 \pi^\frac{1}{2} a}{r} \right)^6 
\left( \dfrac{\im \tilde \alpha}{a_\sigma^2 \lambda} \right)^2
$ 
& $
\dfrac{32 \pi}{3}
\left( \dfrac{\im \tilde \alpha}{a_\sigma^2 \lambda} \right)^2
$
& 1 
  \\ 
\hline 
$\Delta \Gamma_\mathrm{A} $
& $ 
\mp  2 \pi  \left(\dfrac{\lambda_{nk}}{2 \pi r}\right)^4    
\left( 
\dfrac{a_\sigma}{r}
\right)^2
 
\left( 
\dfrac{a_\alpha}{2 \pi^\frac{1}{2} a }
\right)^2
$
& $
\mp
\dfrac{3}{\pi}
\left( \dfrac{a_\alpha}{r} \right)^3
\left( \dfrac{2 \pi^\frac{1}{2} a}{r} \right)^3
$
&  $
\mp
\dfrac{8 \pi}{3}
\left( \dfrac{a_\alpha}{2 \pi^\frac{1}{2} a} \right)^3
$
&$
\mp
\dfrac{1}{2}
\left( \dfrac{a_\alpha}{\im \tilde{\alpha}} \right)^3
\left( \dfrac{a_\sigma}{2 \pi^\frac{1}{2} a} \right)^2
\left( \dfrac{\lambda_{nk}}{4 \pi^\frac{3}{2} a} \right)
$
 \\ 
\hline 
\multicolumn{5}{|c|}{$\lambda_{nk}/2\pi > r > 2 \pi^\frac{1}{2} a \sim a_\sigma  \gtrsim a_\alpha \quad \Rightarrow  \quad 
\Gamma_\mathrm{A,0} > \Gamma_\mathrm{ICD} \gtrsim \Delta\Gamma_\mathrm{A}  \gtrsim \Gamma_\mathrm{s,0} \sim \Delta \Gamma_\mathrm{s} $}
\\
\hline
\end{tabular} 
\caption{
Comparison of different relaxation rates for an excited atom in close proximity to a second atom (via one-photon-exchange). In the presence of a second atom, the one-atom decay rates $\Gamma_\mathrm{s/A}$ gain a contribution $\Delta \Gamma_\mathrm{s/A}$. 
The ratios are defined by the ratio of the length scale: 
transition wavelength $\lambda_{nk}$, atom-separation $r$, Auger-radius $a$, photoionisation radius $\pi a_\sigma^2 = \sigma$ and polarisability radius $a_\alpha ^3 = \tilde{ \alpha}$.
If the involved length scales obey their typical relation to each other, we can sort the rates by their magnitude. For the given grading we excluded the case of $\wkn \approx \omega_i$, which would lead to a significantly high $\im \alpha$ (see Eq.~\eqref{alphawkn}). 
}
\label{tab:ratios}
\end{table*}
The most fundamental relaxation channel of a single atom is spontaneous decay $\Gamma_\text{s}$. In terms of the Green's tensor it is given by \cite{Dung2000}:
\begin{align}
\Gamma_\text{s} &=\frac{2\mu_0}{\hbar} \sum \wkn ^2 \bm d_{nk} \cdot \im \G(\bm r, \bm r, \wkn ) \cdot \bm d_{kn}
\label{channels:spont0}
\end{align}
where the sum runs over all degeneracies. We assume for simplicity that the process is isotropic:
\begin{align}
\Gamma^\iso_\text{s} 
&=\frac{2 \mu_0}{3 \hbar} |\bm d_{nk} |^2 \wkn ^2 \tr \im \G(\bm r, \bm r, \wkn ) 
\end{align}
In free space this gives the well known spontaneous decay rate: 
\begin{align}
\Gamma^\iso_\mathrm{0,s} = \gamma_{nk} 
= \frac{\wkn^3 }{3\pi \varepsilon_0 c^3 \hbar }|\bm d_{nk}|^2
\end{align}
For sufficiently high excitation Auger decay becomes additionally available as a decay channel. For simplicity we assume that the exchange term $m \rightarrow k$ is dipole forbidden and that the process is isotropic. The free space Auger rate is hence given by Eq.~\eqref{augerfreespace}.
Typically once the Auger decay is energetically allowed it is much faster than the spontaneous decay rate by a factor of: 
\begin{align}
\frac{\Gamma_\mathrm{0,A} }{ \Gamma_\mathrm{0,s}}
= \frac{2 \pi^2}{3} \left( \frac{\lambda_{nk}}{4\pi^{3/2} a } \right)^4\times \left( \frac{a_\sigma }{2 \pi^{1/2} a } \right)^2
\label{AvsS}
\end{align}  
where $\lambda_{nk} = 2 \pi c/\wkn$ is the wavelength of the initial transition and we defined the photoionisation radius $\pi a_\sigma^2 = \sigma_m(\wkn)$. 
The photoionisation cross section $\sigma_m(\omega)$ decreases with some order of $\omega$, depending on the orbital quantum number $l$ of state $\ket{m}$ (f.e.~ for an $s$-state $\sigma$ decreases with $\omega^{-11/2}$, for a $p$-state with $\omega^{-15/2}$ \cite{Drukarev2019}) and is usually in the order of $\sim \mathcal{O}(10^{-2}) - \mathcal{O}(10^{1})$ Mb. 
The photoionisation radius $a_\sigma$ is hence in a similar regime as $2 \pi^\frac{1}{2} a\sim \mathcal{O}(10^{-1}) - \mathcal{O}(10^{-1}) $ \AA. 
However for the transition wavelength holds: $\lambda_{nk}/2\pi \gg 2 \pi^\frac{1}{2} a$, which decides the ratio~\eqref{AvsS} in favour of Auger decay. 
The presence of a second atom may influence these rates. Even at distances where the wave function overlap may be neglected the second atom passively manipulates the electromagnetic vacuum and serves as a mediator for the radiative rate as well as the Auger decay rate. 
The mediation by the second atom is governed by the appropriate scattering Green's tensor~\eqref{QED:G1med}. We assume an isotropic polarisability: $\bm \alpha = \alpha \ten{I}$, and introduce the polarisability volume $\tilde \alpha = \alpha /4 \pi \varepsilon_0$. It is given by: 
\begin{align}
 \alpha(\wkn)  
&= 
\frac{2}{3\hbar}\sum_i \frac{\omega_i | \bm d_{in} |^2 }{\omega_i^2 -\wkn^2  + \mi \wkn \gamma_i} 
\end{align}
with resonances at $\omega_i$ with a width of $\gamma_i$. 
Depending on $\wkn$, $\alpha$ can be devided in three different regimes: 
\begin{align}
&\text{For }\wkn \ll \omega_i: & \tilde\alpha \approx \frac{\alpha_0}{4 \pi \varepsilon_0} &\propto \sum_i   \frac{|\bm d_{in}|^2}{\omega_i } 
\nonumber
\\
&\text{For }\wkn = \omega_i: & \tilde\alpha  &\propto \mi \frac{|\bm d_{in}|^2}{\gamma_i} 
\label{alphawkn}
\\
&\text{For }\wkn \gg \omega_i: &\tilde\alpha   &\propto  \sum_i  \left( \frac{\omega_i }{\wkn }\right)^2  \times \frac{|\bm d_{in}|^2}{\omega_i } 
\nonumber
\end{align}
where $\alpha_0$ is the static polarisability and $\tilde\alpha$ is only non-real, when close to a resonance. We will exclude cases $\wkn \ll \omega_i$ from this discussion. 
In the nonretarded regime the rates in presence of a second atom can be given by:
\begin{align}
\Gamma_\text{s}
&\approx
\Gamma_\text{s,0}
\left( 1
+3 \left( \frac{\lambda_{nk}}{2 \pi r}\right)^3   \frac{\im \tilde \alpha}{r^3}
\right)
\\
\Gamma_{ \text{A}}
&=
\Gamma_{0,\text{A}}
\left(
1 
-
\frac{24 \sqrt \pi a^3 \re \tilde \alpha }{ r^6}
+ \frac{216 \pi a^6 | \tilde \alpha|^2 }{ r^{12} }
\right)
\end{align}
Only for $\wkn \approx \omega_i$ the spontaneous decay rate is significantly enhanced compared to the Auger rate. 
The magnitude of $\tilde \alpha$ on a resonance is determined by its line width $\gamma_i$ and can be of several orders of magnitude. For transition energies greater than the resonances in the mediator $\tilde \alpha$ decreases with $\wkn$. In this energy regime typically $\tilde \alpha \leq \mathcal{O}(10^{-1})$ \AA $^3$. We define a length scale $a_\alpha$ for the polarisability volume $\tilde \alpha = \pm  a_\alpha^3$. With this we find for the Auger rate in first order:
\begin{align}
\Gamma_{ \text{A}}
&\approx
\Gamma_{0,\text{A}}
\left[
1 
\mp 4
\left( \frac{2 \sqrt{\pi} a}{ r}\right)^3
\times
\left( \frac{a_\alpha }{ r}\right)^3
\right]
\end{align}


By introducing a second atom into the system we also open up another relaxation channel, i.e.~ICD. The isotropic free-space ICD rate in the nonretarded limit is given by Eq.~\eqref{GammaICDiso0NR}. 
The ratio between ICD and Auger rate is given by: 
\begin{align}
\frac{\Gamma_\text{A } }{\Gamma_{\text{ICD} }}
&\approx
\frac{\Gamma_\text{A,0} }{\Gamma_{\text{ICD}}}
+
\frac{\Delta \Gamma_\text{A} }{\Gamma_{\text{ICD}}}
\\
\frac{\Delta \Gamma_\text{A} }{\Gamma_{\text{ICD}}}
&=
\mp  \frac{32 \pi^2}{9} 
\left( \frac{ a_\alpha}{ 2 \sqrt{\pi} a}\right)^3
\\
\frac{\Gamma_\text{A,0} }{\Gamma_{\text{ICD} }}
&=
\frac{8 \pi^2 }{9 } \left( \frac{ r }{2 \sqrt{\pi} a } \right)^6
\end{align}
The free space ratio is usually much larger than one as a result of $r/a >1$. The different ratios are given in a compact form in table \ref{tab:ratios}, together with an estimation of the rates proportion, where we assumed that $\wkn \gg \omega_i$, i.e.~that $\im \alpha \ll 1$.
\section{Impact of a surface onto decay channels}
\label{sec:surface}
\begin{figure}[b]
\includegraphics[width = 0.8 \linewidth]{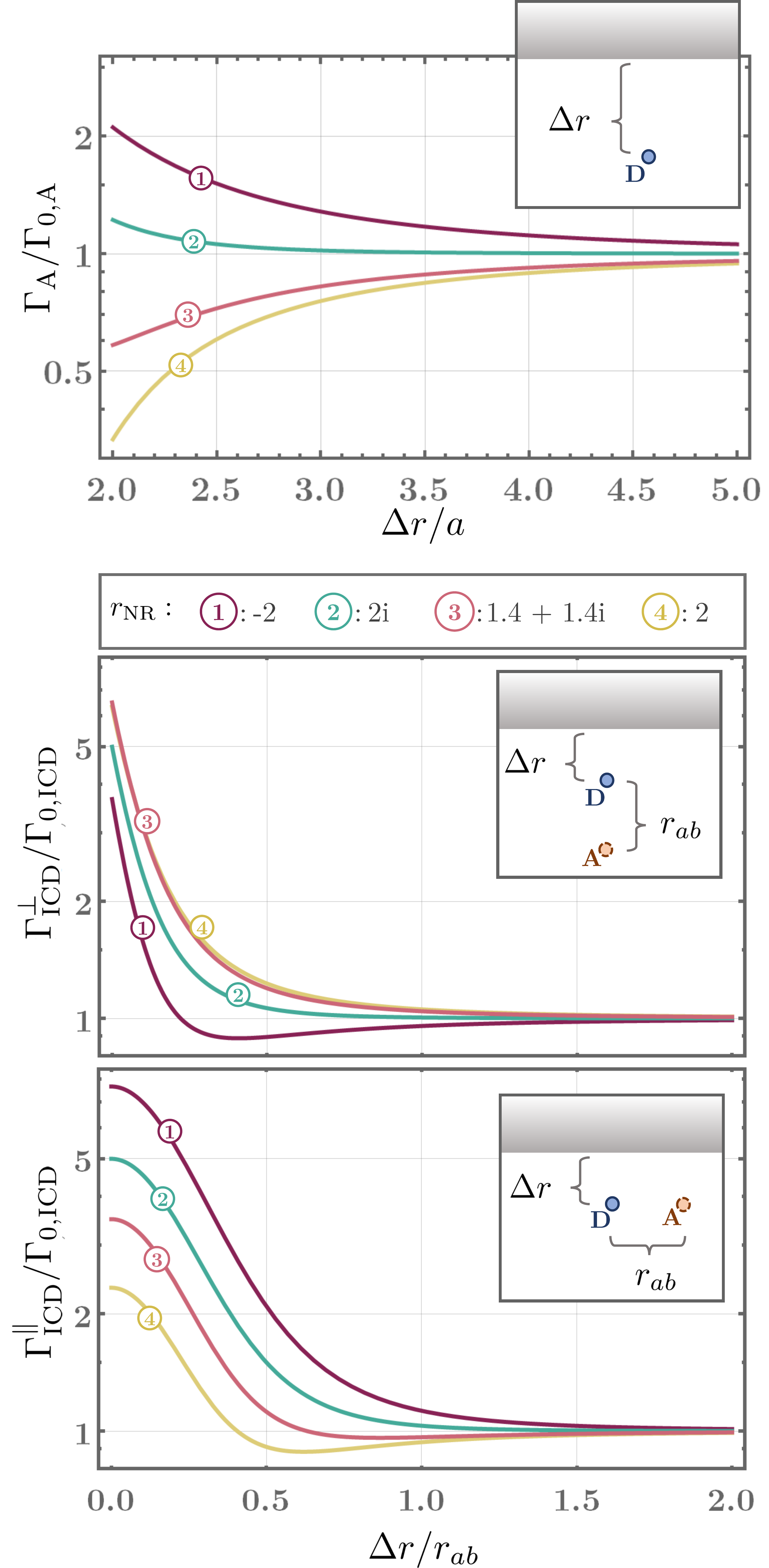}
\caption{The relative Auger rate $\Gamma_\text{A}/ \Gamma_{0,\text{A}}$ and the relative ICD rate $\Gamma_\text{ICD}/ \Gamma_{0,\text{ICD}}$ close to a surface as a function of surface distance $\Delta r$. For ICD two geometries are presented, once where the donor--acceptor-separation $\bm r_{ab}$ is perpendicular to the surface ( $\Gamma_\text{ICD}^\perp$ ) and one where $\bm r _{ab}$ is parallel to the surface ( $\Gamma_\text{ICD}^\parallel$ ). 
Each rate is given for four different reflection coefficients $r_\mathrm{NR} \in \lbrace
-2,2 \mi, 1.4 +  1.4\mi  ,2 \rbrace$ indicated in the curves by their respective index, see Table \ref{tab:eps&n}.
}
\label{fig:icd_auger_iso}
\end{figure}

The propagation of the process-mediating excitation can be influenced via macroscopic bodies in both, ICD and Auger decay. 
For the discussion in this paper we limit ourselves to cases were the relaxing electron transitions from the same energy level for both processes, Auger and ICD. In this case $\wkn$ as well as $\gamma_{nk}$ are the same for both processes, while the photoionisation cross section $\sigma_m( \wkn)$ differs for the two rates. The frequency $\wkn$ determines which materials show the largest effect onto the processes. 
For non-cavity-like geometries the nonretarded regime of very close distances $\Delta r$ to the surface achieve the strongest effects. The scattering Green's tensor can then be approximated by its nonretarded limit and is given by Eq.~\eqref{QED:G1surf}.
We present the impact of a surface for different complex values of 
$r_\mathrm{NR} \in \lbrace -2 , 2 \mi , 1.4 + 1.4 \mi ,2 \rbrace$ ($|r_\mathrm{NR}| = 2$). In Table~\ref{tab:eps&n} the respective permittivity and complex refraction index $n_r = \sqrt{\varepsilon}$ for these values are given.
\begin{table}
  \begin{center}
    \begin{tabular}{l|c|c|c} 
    \hline
		  index  
		& $r_\mathrm{NR}$ 
		& $\varepsilon$ 
		& $n_r$ 
	\\   \hline
   		1 \,
   		&$-2 	$			
   		& $ - 0.33$ 		
   		& $0.58$i \, 
   	\\
   		2 \,	
   		&2$\mi$ \, 			
   		&\, $ -0.60+0.80$i\, 
   		& \, $0.45 + 0.90$i \,  
   	\\
  		3\,
  		&\, $1.41 +1.41$i      \,
  		&\, $-1.38 + 1.30$i    \,
  		&\,  $0.51 + 1.28$i  \, 
  	\\
   		4 \, 	
   		&2 					
   		&$ - 3$					
   		&$1.73$i 
   	\\
    \end{tabular}
    \caption{Chosen values for the material parameters at $\omega = \wkn$. The parameters are related by: $r_\mathrm{NR} = (\varepsilon -1)/(\varepsilon +1)$ and $n_r = \sqrt{\varepsilon}$.
    }
    \label{tab:eps&n}
  \end{center}
\end{table}
The Green's tensor $\G_{\mathrm{A}} $ for the Auger decay and $\G_{\mathrm{ICD}} $ for ICD in the nonretarded limit are given by: 
\begin{align}
\G_\mathrm{ICD}
&= \G^{(0)} (\bm r_b, \bm r_a, \wkn )+ \G^{(1)}_\text{surface} ( \bm r_b, \bm r_a,\wkn )
\\
\G_{\mathrm{A}}
&= \G^{(0)}_{\mathrm{A}} (\wkn )+ \G^{(1)}_\text{surface} ( \bm r_a , \bm r_a ,\wkn ),
\end{align}
where $\bm r_a$, $\bm r_b$ are the donor and acceptor positions, respectively.
We assume for simplicity that the exchange term  $m \rightarrow k$ is dipole-forbidden. 
Even in the isotropic case there exists a preferred orientation in ICD, given by the separation vector from donor to acceptor. Depending on its relative orientation to the surface the effect onto the process varies, the two extremes being perpendicular or parallel to the surface, see the inset schemes in Fig.~\ref{fig:icd_auger_iso}. 
The rates are then given by:
\begin{align}
\Gamma^{\mathrm{(iso)}}_\mathrm{A}
&= 
\Gamma^{\mathrm{(iso)}}_{0,A} 	\left( 
1 
-	\frac{2 \sqrt{\pi } \re\left[ r_\mathrm{NR}\right] a^3}{\Delta r^3} 
+ 	\frac{9 \pi  | r_\mathrm{NR}|^2 a^6 }{8 \Delta r^6}
  				\right)
\end{align}
\begin{align}
\Gamma^{\mathrm{(iso)}\parallel}_\mathrm{ICD}
&= 
\Gamma^{\mathrm{(iso)}}_\mathrm{0,ICD}
\Bigg(
1
-
\frac{ \re\left[ r_\mathrm{NR}\right] \left(\Delta r^2/r_{ab}^2+4\right)}{3 \left(\Delta r^2/ r ^2_{ab}+1\right)^{5/2}} 
\nonumber
\\
& \qquad\qquad\qquad\qquad\qquad
+
\frac{| r_\mathrm{NR}|^2}{\left(\Delta r^2/r_{ab}^2+1\right)^3}
\Bigg)
\\
\Gamma^{\mathrm{(iso)}\perp}_\mathrm{ICD}
&= 
\Gamma^{\mathrm{(iso)}}_\mathrm{0,ICD}
\left( 
1
+
\frac{ 2 \re\left[ r_\mathrm{NR}\right] r_{ab}^3}{3 \left(2  \Delta r+r_{ab}\right)^3}
+
\frac{| r_\mathrm{NR}|^2 r_{ab}^6 }{\left(2 \Delta r+ r_{ab}\right)^6}
\right)
\end{align}
In Fig.~\ref{fig:icd_auger_iso} these rates are plotted relatively to their free-space rate for the four different permittivities given in Table~\ref{tab:eps&n}.

In this regime the length scale that determines how much the surface influences the process is the atom-separation $r_{ab}$ in ICD, while in Auger this length scale is given by the Auger-radius $a$. The Auger-radius can be determined via known free-space Auger rates or be roughly estimated by using Slater rules of the vacancy orbital \citep{Franz2021} and is of the order of the Bohr radius $a_0 \approx  0.5$\,\AA. Therefore $a \ll r_{ab}$, as a consequence there is a large range of distances $\Delta r$, at which the Auger rate is effectively the free-space rate, while the ICD-rate is influenced. For the chosen permittivities the nonretarded effect of the surface vanishes in case of ICD for separations $\Delta r > 2 r_{ab}$, while the effect onto the Auger rate vanishes for separations $\Delta r > 4 a$. In the nonretarded limit the impact of the surface is comparable to that of an mirrored acceptor-dipole. For same donor distances $\Delta r$ the distance to the mirrored acceptor dipole in the perpendicular case is larger than for the parallel case. The parallel geometry profits from propagation mediation by surface waves.
The two real reflection coefficients $r_\mathrm{NR} \in \lbrace -2, 2 \rbrace$ give the most different behaviour per process and geometry. Every curve belonging to a reflection coefficient with $|\rnr | = 2$  is between these two extremes. For larger values $|\rnr|$ the respective impact of the surface onto the rate would be amplified.
%

\begin{figure}
\includegraphics[width = 0.8 \linewidth]{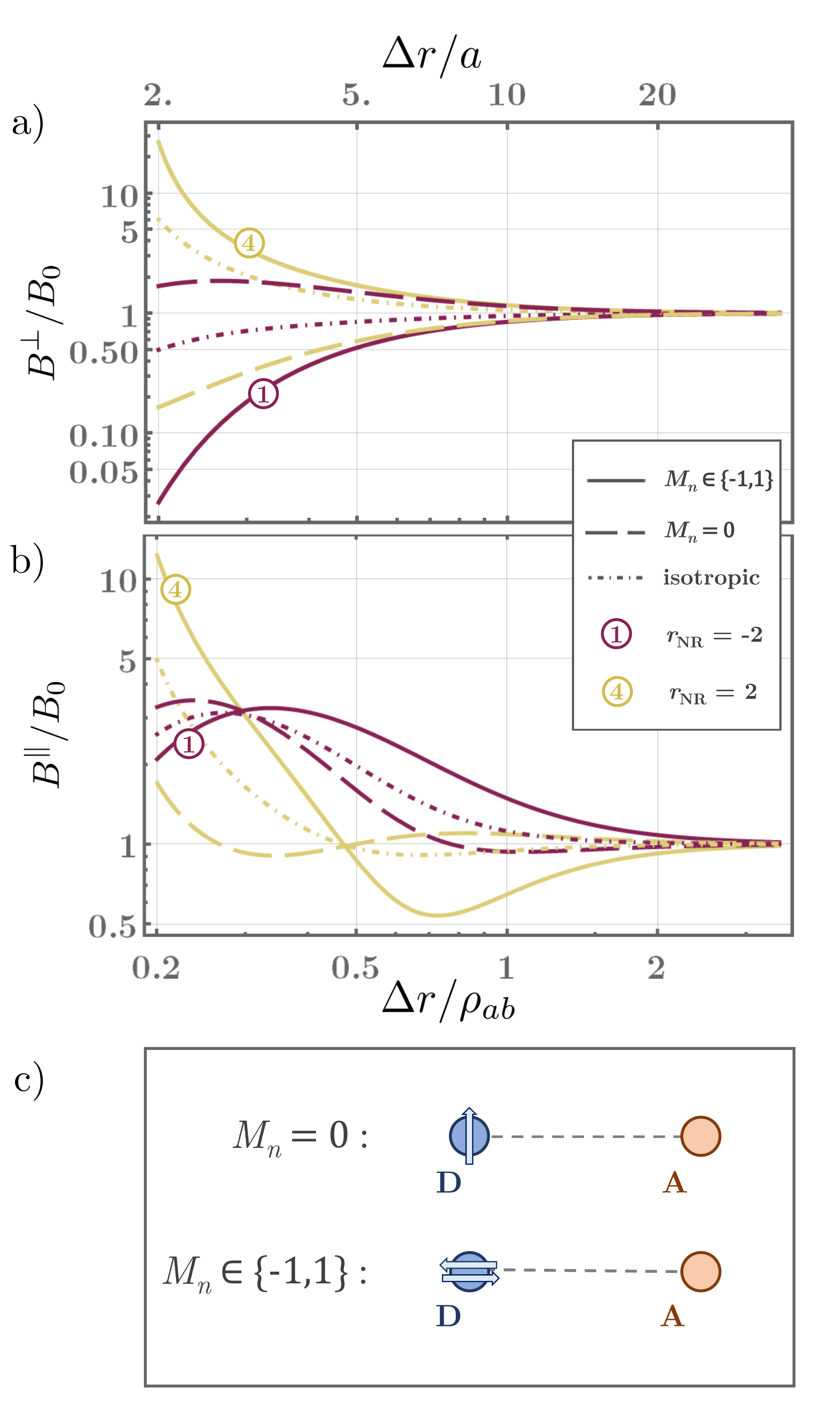}
\caption{The relative branching ratio $B /B_0$ ( see Eq.~\ref{branching ratio}) close to a surface for the perpendicular and parallel ICD-geometries (see Fig.~\ref{fig:icd_auger_iso}) for the two extreme complex phases of the reflection coefficient $r_\mathrm{NR,1(4)} = -2 \,(+2)$. Additionally to the isotropic case, we also considered specific orientations of the transition dipole as a consequence of specific angular momentum projections $M_n$ of the initial state $\ket{n}$. For this plot we chose the characteristic length scale ratio of the two processes to be $r_{ab} /a= 10$.
}
\label{fig:icd_auger_B}
\end{figure}

Since ICD is often studied inside of a dimer or larger molecule, the sum in \eqref{GammaICD1} over the involved transition dipoles $\bm{d}_{nk} = \braket{k|\bm{\hat{d}}|n}$, with $\ket{n} =\ket{E_n,L_n,M_n}$ is not necessarily isotropic. If we introduce transition dipole orientations, we find that specific orientations are stronger influenced by a simple close-by surface than others, depending on the geometry. 
We illustrate this in the example, where the initial $\ket{n}$-state of the donor either only involves 
angular momentum $\bm L_n$ that are either parallel ($M_n \in \lbrace  -L_n,L_n\rbrace$ ) or perpendicular ($M_n = 0$ ) to the quantisation axes, i.e.\,the separation axes between donor and acceptor, see Fig.~\ref{fig:icd_auger_B}c). 
We define the ratio between ICD and Auger rate as branching ratio:
\begin{align}
B = \Gamma_{\mathrm{ICD}}/\Gamma_{\mathrm{A}}
\label{branching ratio}
\end{align}
and the free space branching ratio $B_0 = \Gamma_{0,\mathrm{ICD}}/\Gamma_{0,\mathrm{A}}$, that is constant in $\Delta r$. 
The larger the branching ratio the faster is ICD compared to Auger. The branching ratio itself depends on the ratio of the photoionisation cross sections $\sigma^{(\mathrm{ICD})}_m(\wkn)/\sigma^{(\mathrm{A})}_m(\wkn)$, which can be of several orders of magnitude. The impact of the environment onto the branching ratio $B/B_0$ however depends in the nonretarded limit only on the surface properties and the geometry, including the relation of $r_{ab}$ and $a$. In Fig.~\ref{fig:icd_auger_B} the branching ratio is given for $a = 10r_{ab}$ compared to the free space branching ratio. 
Both extreme complex phases of the reflection coefficient $\rnr \in \lbrace -2,2 \rbrace$ are presented. 
The introduced dipole orientations are calculated separately as well as the isotropic case. For the perpendicular geometry (Fig.~\ref{fig:icd_auger_B}a) the branching ratio $B^{\perp}$ shows a simple behaviour as function of the surface distance $\Delta r$: 
For $M_n \in \lbrace -1 ,1 \rbrace$ as well as for the isotropic initial state the branching ratio is shifted in favour of ICD for all distances for $\rnr  > 0$, while for $M_n = 0$ the ICD-Auger-ratio is enhanced for $\rnr < 0$.
For the parallel geometry a negative reflection coefficient $\rnr <0$ leads in all cases to an enhanced ICD-Auger-ratio $B^\parallel/B_0 > 0$. A positive reflection coefficient can shift the branching ratio in either direction and achieves in very short surface distances an higher enhancement than the negative reflection coefficient.

\begin{figure*} 
\includegraphics[width = \textwidth]{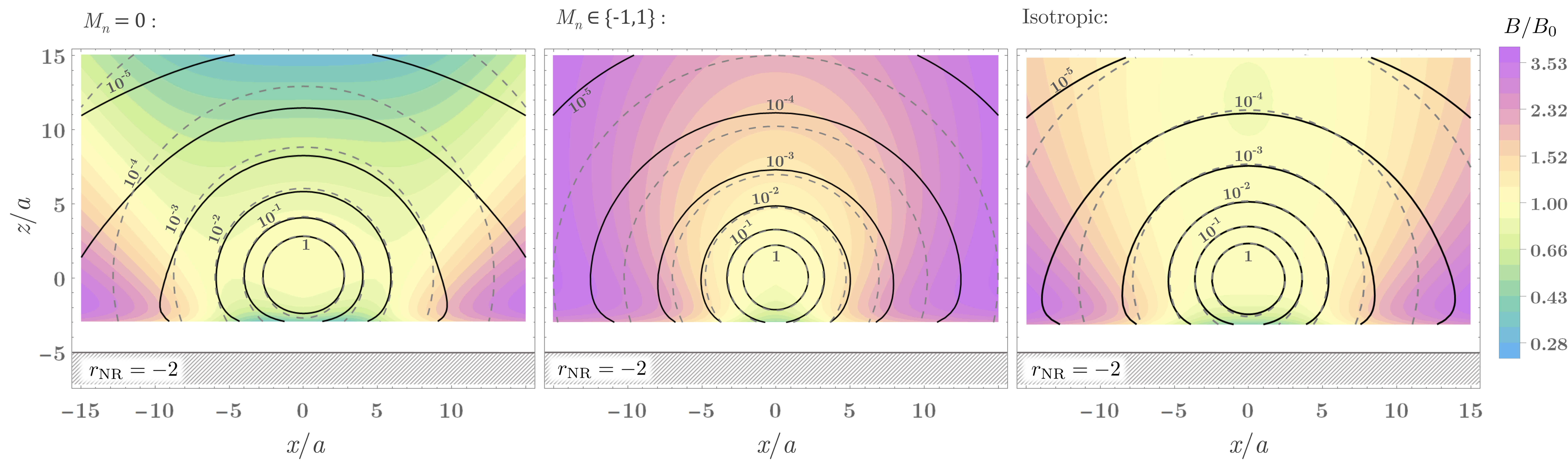}
\caption{Contour plots for the two different non-isotropic cases ( $M_n = 0$ and $M_n \in \lbrace -1,1\rbrace$) as well as the isotropic case for a surface reflection coefficient $r_\mathrm{NR} = -2$. 
The acceptor atoms position is fixed at the origin, while we vary the position of the donor.
The black contours show the absolute branching ratio in terms of the cross section ratio $B / (\sigma_\mathrm{ICD}/\sigma_\mathrm{A})$ in the presence of the surface, while the dashed contours show the respective value for the branching ratio in free space $B_0 / (\sigma_\mathrm{ICD}/\sigma_\mathrm{A})$. The color plot shows the difference between the branching ratio with and without the surface $B/B_0$. 
}
\label{fig:contours}
\end{figure*}

In Fig.~\ref{fig:contours} the distance between the two atoms is not fixed. Instead we fix the position of the acceptor atom at $\Delta r = 5 a$, where $a$ is again the Auger-radius and hence the fundamental scale for Auger decay. The donor atoms position is varied. 
The contours give the branching ratio~\eqref{branching ratio} between ICD and Auger for a donor at the respective position in terms of the photoionisation cross section ratio , if f.e.~the ratio of the respective photoionisation cross sections gives $\sigma^{(\mathrm{ICD})}_m /\sigma^{(\mathrm{A})}_m = 100$ then the contour at $B/ (\sigma^{(\mathrm{ICD})}_m /\sigma^{(\mathrm{A})}_m ) = 0.01$ gives the donor position at which Auger and ICD are equally fast. 
The larger the donor--acceptor distance the more the surface can enhance ICD over Auger. However the larger the donor--acceptor distance the lower is the free space branching ratio $B_0$. A system  were the photoionsiation cross section of the ICD process is much larger than the one of Auger gives a more preferable initial condition. This would lead to a higher branching ratio at larger donor--acceptor-distances which can be enhanced more easily by an appropriate surface. The impact of the surface becomes stronger with larger $|r_\text{NR}|$. The strongest effect can be achieved if the involved transition dipoles are parallel to the surface. 

In general if the reflection coefficient is around the magnitude of 1, the distances at which a surface shows significant influence onto the rates are smaller or equal to the donor--acceptor separation $r_{ab}$. At such distances there may occur additional effects that one has to account for. The discussed results should still serve as a good approximation. For a given material it is also possible to account for local field effects by modelling the single constituents of the material via their appropriate polarisability tensor. The polarisability of single atoms can be related to the permittivity via the Clausius–Mossotti relation.

\section{Cavity discussion}
\label{sec:cavity}
The Green's tensor formalism exploited so far can also be used to calculate the respective rates in a cavity.
For a explicit calculation the specific cavity must be considered and the Green's tensor can be calculated via numerical methods, only a perfect spherical cavity leads to an analytic solution. However it is possible to use our formalism to estimate the effect of a cavity onto ICD and Auger depending on its $Q$-factor. 

The $Q$-factor of a cavity is defined via the relation between the spontaneous decay rate in free space $\Gamma_{s,0}$ and the one enhanced by the cavity $\Gamma_{s}$:
\begin{align}
Q &= \frac{\Gamma_{s}}{s \Gamma_{s,0}}, 
\qquad
\text{with:} \,\,
s = \frac{3 \lambda ^3}{4 \pi^2 V}
\\
\Rightarrow
sQ -1  &=\frac  {| \im \G^{(1)} (\bm r_a, \bm r_a) |}{  |\im \G^{(0)} (\bm r_a, \bm r_a) |},
\label{cav:Qdef} 
\end{align}
where $\bm r_a$ is the position of the atom.
Cavity QED usually assumes the opposite limit of the one used so far, namely the retarded limit in the surface-system-distance $\Delta r \wkn /c \gg 1$. In the retarded limit the scattering Green's tensor describes propagating waves and we may approximate: $|\G^{(1)} | \approx |\im \G^{(1)} |$. We assume that a system undergoing ICD in the cavity has a donor--acceptor-separation $r_{ab}$ that is much smaller than the surface-system-separation $\Delta r$. The scattering Green's tensor for ICD, can hence be approximated by: 
$\G^{(1)} (\bm r_b, \bm r_a ) \approx \G^{(1)} (\bm r_b = \bm r_a)$ which is the same scattering Green's tensor as for spontaneous decay. We also use that the non retarded bulk Green's tensor $\G^{(0)}(r_a,r_b)$ and the regularised bulk $\tilde \G $ are real.
To summarize: 
\begin{subequations}
\begin{alignat}{3}
r_{ab} 
&
\ll \Delta r 
&
\Rightarrow 
&
&
\G^{(1)}(r_b,r_a) 
&\approx \G^{(1)}(r_a,r_a)
\label{cav:G1ab}
\\
\wkn \Delta r /c 
&
\gg 1  
&\Rightarrow  
&
&
\G^{(1)}(r_a,r_a) 
&\approx \im \G^{(1)}(r_a,r_a) 
\nonumber
\\
& 
&& 
&&\approx \re \G^{(1)}(r_a,r_a)
\label{cav:imG1}
\\
\wkn r_{ab} /c 
&
\ll 1
&\Rightarrow  
&
&
\G^{(0)}(r_a,r_b) 
&\approx \re \G^{(0)} (r_a,r_b)
\label{cav:reG0}
\\
\wkn a /c 
&
\ll 1
&\Rightarrow  
&
& \tilde{ \G}^{(0)}&=\re \tilde{ \G}^{(0)} 
\label{cav:reG0a}
\end{alignat}
\end{subequations}
We additionally define the ratio between the imaginary part of the Green's tensor in free space as needed for the calculation of spontaneous decay (see \eqref{channels:spont0}) and the real part of the Green's tensor for ICD and Auger, respectively:
\begin{subequations}
\begin{align}
b_\mathrm{icd} &= \frac{ | \im \G^{(0)}(r_a,r_a) | }{ | \re \G^{(0)}(r_b,r_a)| } 
= \frac{\wkn^3  r_{ab}^3  }{3 c^3} \ll 1.
\label{cav:bicd}
\\
b_\mathrm{a} &= \frac{| \im \G^{(0)} (r_a,r_a) | }{|  \re \tilde{\G}^{(0)} | } 
= 4 \sqrt{\pi}
\frac{\wkn ^3 a ^3}{c^3} \ll 1
\label{cav:ba}
\end{align}
\end{subequations}
With these approximations we can estimate the maximum possible enhancement for ICD in a cavity: 
\begin{align}
\Gamma_\text{ICD}^{\text{(cav)}} 
&= \Gamma_\text{ICD,0}
\bigg( 
\frac{| \G^{(0)}  + \G^{(1)}  |^2}
{| \G^{(0)} |^2}
\bigg)
\nonumber
\\
&=
\Gamma_\text{ICD,0}
\bigg( 
1 
+ 
2 
\frac{ 
\re \G^{(1)}  
}
{\re \G^{(0)} }
+ 
\left[
\frac{ \im \G^{(1)}}
{\re \G^{(0)} } \right]^2
\bigg)
\nonumber
\\
&=
\Gamma_\text{ICD,0}
\bigg( 
1 
+ 
2 
( s Q - 1) b_\mathrm{icd} 
+
( s Q - 1)^2 b_\mathrm{icd}^2 
\bigg)
\nonumber
\\
&\approx 
\Gamma_\text{ICD,0}
\left( 
1 + 2 s b_\mathrm{icd} Q + s^2 b_\mathrm{icd} ^2 Q^2 
\right)
\end{align}
where we assumed, that $s b Q \sim 1$ and omitted the position arguments for simplicity. Similarly, by using \eqref{cav:reG0a} and \eqref{cav:ba} we find for the maximum enhanced Auger decay rate in a cavity:
\begin{align}
\Gamma_\text{A}^{\text{(cav)}} 
&\approx \Gamma_\text{A,0}
\left( 
1 + 2 s b_\mathrm{a}  Q + s^2 b_\mathrm{a} ^2 Q^2 
\right)
\end{align}
To achieve maximum difference between ICD and Auger decay, donor and acceptor should be placed onto the appropriate phases of the standing electromagnetic wave inside the cavity.
\section{Application to He-Ne-Dimer}
\label{sec:HeNe}
\begin{figure}
\includegraphics[width =  \linewidth]{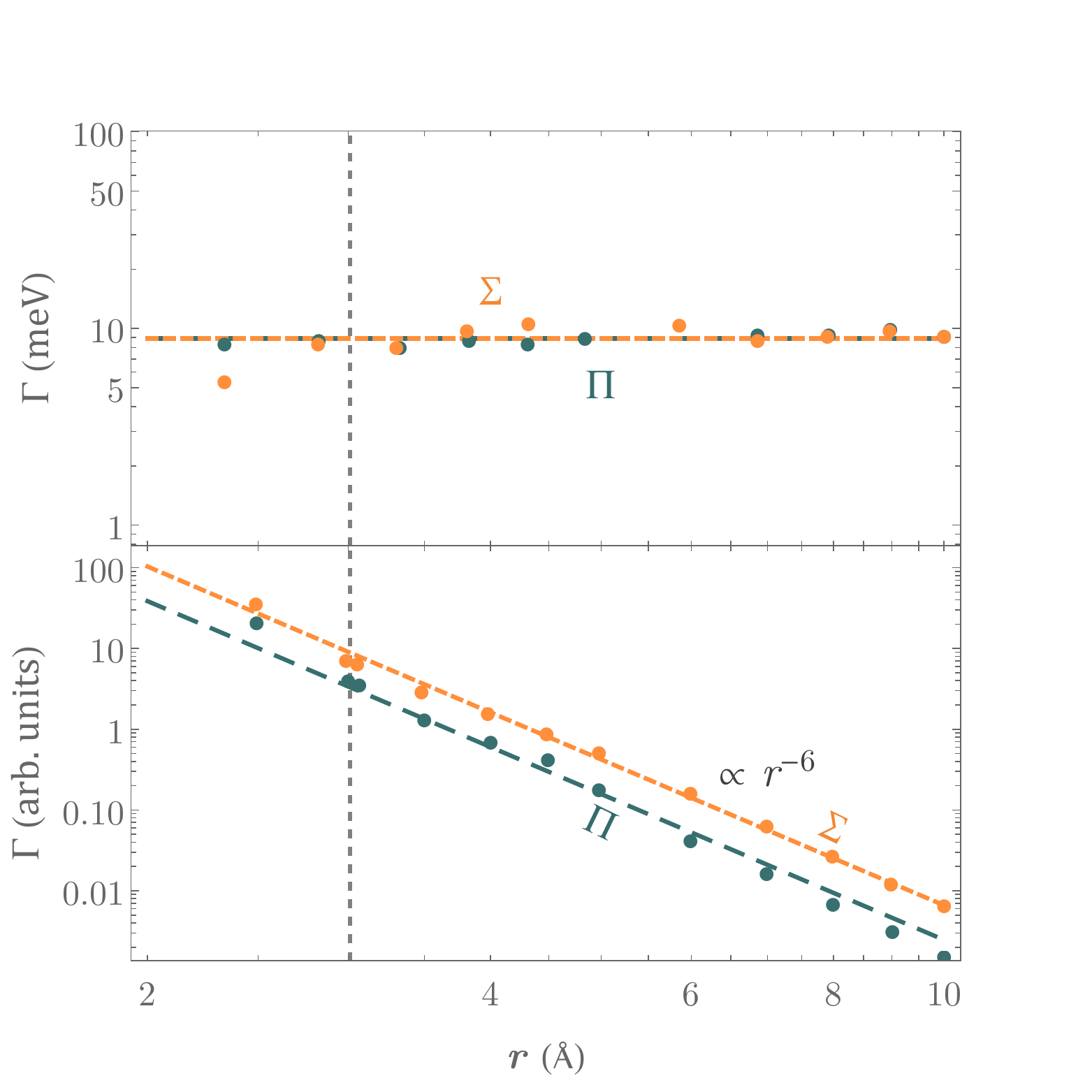}
\caption{
The free space ICD and Auger decay rate calculated by Jabbari et al.~\cite{Jabbari2020} for different dimer separations $r_{ab}$. Even at very small distances the ICD rates decrease with $r_{ab}^{-6}$, as illustrated by the dashed lines and the Auger rates stay independent of $r_{ab}$. This implies that wave function overlap between neon and helium do not play a significant role for the rates, instead we may use single atom data to approximate the rates in our framework. The gray vertical line at $r_{ab}=3.01$~\AA\,marks the equilibrium distance between He and Ne in the dimers ground state.
}
\label{fig:r^-6}
\end{figure}

\begin{figure}
\includegraphics[width = 0.8 \linewidth]{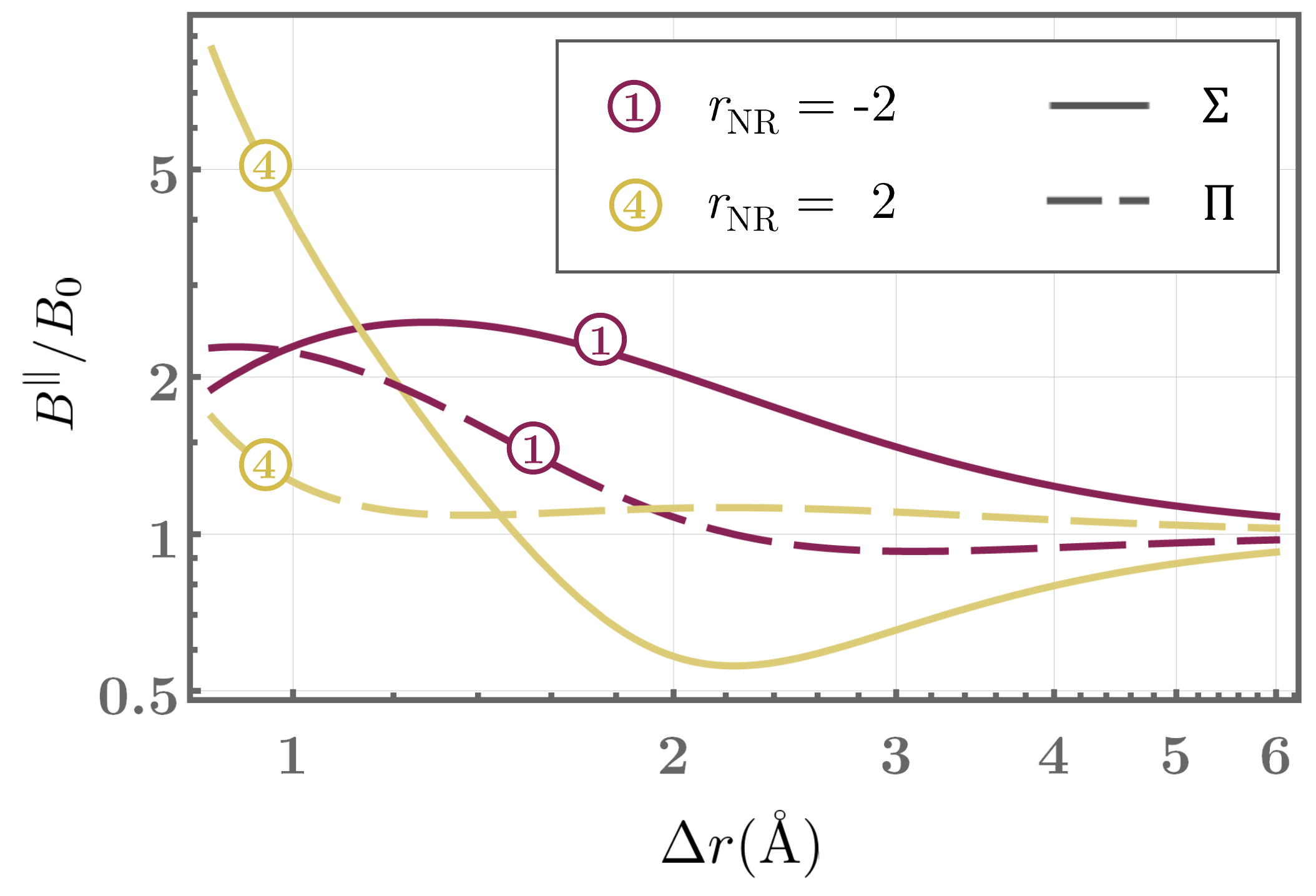}
\caption{The relative branching ratio when placing a HeNe-dimer with a donor--acceptor-distance of 3.01\AA~in front of a surface with a reflection coefficient of $r_\mathrm{NR} \in \lbrace -2,2\rbrace$. The curves are labelled by the respective index of their reflection coefficient, see Table \ref{tab:eps&n}. 
The HeNe-dimer can either be in a $\Pi$- or $\Sigma$-state, leading to different branching ratios in the presence of a surface. The chosen initial state of the doubly-excited helium corresponds to the 23sp+ single-atom state.
The characteristic length-scale for the Auger decay in 23sp+ helium is $a = 0.431$\AA . We chose $\bm r_{ab}$ to be parallel to the surface.
}
\label{fig:B_HeNe}
\end{figure}

An example of a system, where Auger decay and ICD compete each other is the HeNe-dimer, where helium is doubly excited \citep{Jabbari2020}. In their ground state (before  excitation) the two atoms have a separation of $r_{ab} =3.01$ \AA\citep{Jabbari2020}. The dimer exists in two possible molecular state $\Sigma$ and $\Pi$.
For large separation differences the molecule state maps to the product of the single atom ground state of neon and the $M_\Sigma = 0$ and $M_\Pi \in \lbrace-1,1\rbrace$ state for helium, respectively. 
When exciting helium there are several possible doubly excited states. The two dominating ones are 2s2p and 23sp+, with $\ket{23sp+}= 2^{-1/2}\left(\ket{2p3s}+ \ket{2s3p}\right)$. For the dimer consisting of 23sp+ helium the free space rates of Auger and ICD are robust against wavefunction overlap corrections that are not goverend by our theory, which can be seen at the $r^{-6}$-behaviour of the numerically calculated rates by Jabbari et al.~\citep{Jabbari2020} as a function of the dimer separation, see Fig.~\ref{fig:r^-6}.
To apply our formalism we need to determine the involved single atom properties $\wkn$, $\gamma_{nk}$ and $\sigma_m$ of the respective process:
\begin{align}
\text{ICD:} \qquad\qquad
\text{He}^{**}\text{(23sp+)} 
\quad
&\underset{\gamma_{n\!k} }{\longrightarrow}
\quad \text{He}^{*}\text{(1s3s)} + \hbar \wkn
\nonumber
\\
\text{Ne} + \hbar \wkn
\quad
&\underset{\sigma^{\text{ICD}}_m  }{\longrightarrow}
\quad \text{Ne}^+ + \text{e}^-
\nonumber
\\
\text{Auger:} \qquad\qquad
\text{He}^{**}\text{(23sp+)} 
\quad
&\underset{\gamma_{n\!k}}{\longrightarrow}
\quad \text{He}^{*}\text{(1s3s)} + \hbar \wkn
\nonumber
\\
\text{He}^*\text{(1s3s)} + \hbar \wkn
\quad
&\underset{\sigma^{\text{A}}_m}{\longrightarrow}
\quad \text{He}^+ + \text{e}^-
\nonumber
\end{align}
They are given by: $\wkn = 40.94$ eV \citep{Jabbari2020}, $\sigma_m^{\text{ICD}} = 9.28$ Mb \citep{Verner1995}, $\sigma_m^\text{A} =0.35  $  Mb \citep{NicoPrivate}, $\gamma_{nk} = 5.65 \times 10^9$ s$^{-1}$ \citep{Liu2001}. With this we can determine the Auger-radius for Auger decay: $a = 0.457$ \AA. 
Together with the equilibrium distance for $r_{ab}$, we find a ratio of $r_{ab} /a = 6.58$. With this ratio we find the strongest effect for a parallel geometry and the $\Pi$-dimer, see Fig.~\ref{fig:B_HeNe}. For a negative reflection coefficient $r_\text{NR} = -2$ one can reach an enhancement of the branching ratio $B/B_0 = 2$ at a 2~\AA\,  distance, while a positive reflection coefficient shifts the ratio in favour of Auger to be $B/B_0 \approx 1/2$ at the same distance. At very close distances the ICD process can be enhanced for $r_\text{NR}=-2$ significantly, however as mentioned before it would be appropriate to exploit local field methods in this close realm that resolve the structure of the specific material by using its density and the polarisability of its constituents. 
For arbitrary donor--acceptor distances in the presence of the considered surface with an nonretarded reflection coefficient $r_\mathrm{NR} = -2$ we can use the known photoionisation cross section ratio $\sigma_m^\mathrm{ICD}/\sigma_m^\mathrm{A} = 26.76$ to label the contours in Fig.~\ref{fig:contours} accordingly, f.e.~when placing the donor at the $B/(\sigma_m^\mathrm{ICD}/\sigma_m^\mathrm{A} ) = 10^{-2}$-contour we expect a branching ratio of $B = 0.27$, which means the ICD rate would be roughly a quarter of the Auger rate. 

The polarisability $\alpha_\text{Ne}$ of neon itself is negligible at 43.84~eV ( $\alpha_\text{Ne}( 43.84$~eV$) \sim \mathcal{O}(10^{-6})$) and therefore neither mediates the radiative nor Auger decay \citep{CXRO}.

In a cavity we can determine the necessary $Q$-factor for each process by determining the defined $b$-factors:
\begin{align}
b_\mathrm{icd} 
&=  7.99 \times 10^{-5}, 
\qquad
b_\mathrm{a} 
= 5.78 \times 10^{-6}.
\end{align}
To achieve a significant effect via a cavity its $Q$-factor for an transition frequency of $\wkn = 40.94$ eV hence needs to be at least of the order of $sQ \sim \mathcal{O}(10^4)$ and $\sim \mathcal{O}(10^5)$ for ICD and Auger decay, respectively.

\section{Conclusions}
Interatomic Coulombic decay (ICD) and Auger decay rates were derived as special cases of electron--electron scattering. By applying the dipole-approximation to the ICD rate and a recently proposed finite electron-cloud approximation to the Auger decay rate, we were able to find analytic expressions for both. 
We determined the atomic properties that one needs to consider to predict the ratio of the various rates in free space and by working in the framework of macroscopic quantum electrodynamics we are able to consider macroscopic bodies via a classical Green's tensor, which properties are well studied for many different systems.
\\ \indent
As an example, we introduced a simple close-by surface to an excited two-atom system and could show that, depending on its nonretarded reflection coefficient, even this simple set-up can lead to a change of the rate ratio in favour of either ICD or Auger decay.
We have shown that in the nonretarded limit the length scale for distances at which the surface can influence the process is given by the donor--acceptor distance for ICD and by the Auger-radius for Auger decay. The effect scales with the absolute value of the complex nonretarded reflection coefficient.
In addition, we considered non-isotropic transitions and showed that specific transition-dipole orientations can lead to enhanced effects. 
Finally, we considered a cavity via its $Q$-factor and estimated its maximum enhancement onto ICD and Auger decay. While the surface study exploits the nonretarded regime of close distances to the macroscopic body, the cavity estimation needs to be considered in the retarded regime. We related the enhancement inside of the cavity to the characteristic length scales of each process compared to the wavelength of the initial de-excitation. 
\\ \indent
The provided general expressions and graphs can be applied to different specific systems. 
When applied to the example of a HeNe-dimer we showed that it is easier to change the ratio in favour of ICD when considering the $\Pi$-state of the dimer than the $\Sigma$-state, which is a direct consequence of the transition dipole orientation relative to the surface. 
\\ \indent
The presented methods can be used to predict the effect of specific surfaces, additional atoms in the system, or a cavity onto the excitation propagation in ICD and Auger decay. 
Additional effects onto the electronic density of states or even expected level shifts inside the atoms can be taken into account by appropriate replacement of the photoionisation cross section and spontaneous decay rate in the respective rate expression. For a specific surface where one would expect to see the presented effects, the calculation can be improved in the limit of very close distances by resolving the material constituents via their polarisability. 
With the presented derivation it is also possible to go beyond dipole approximation and perform \textit{ab initio} calculations while taking the scattering onto macroscopic bodies into account via the appropriate Green's tensor.
\acknowledgments{
The authors thank R.~Bennett, A.~Burkert, L.~S.~Cederbaum, K.~Gokhberg, T.~Janka, M.~Kowalewski, D.~Lentrodt, N.~Sisourat for discussions. This work was
supported by the German Research Foundation (DFG,
Grants No. BU 1803/3-1 and No. GRK 2079/1).
}

\bibliography{library_manual_fix}
\end{document}